\documentstyle[12pt,epsf]{article}

\def\NPB{{\em Nucl. Phys.} B}
\def\PLB{{\em Phys. Lett.}  B}

\def\PRD{{\em Phys. Rev.} D}

\newcommand{\nwc}{\newcommand}
\nwc{\cl}  {\clubsuit}

\nwc{\hyp} {\hyphenation}
\nwc{\be}  {\begin{equation}}
\nwc{\ee}  {\end{equation}}
\nwc{\ba}  {\begin{array}}
\nwc{\ea}  {\end{array}}
\nwc{\bdm} {\begin{displaymath}}
\nwc{\edm} {\end{displaymath}}
\nwc{\bea} {\be\ba{rcl}}
\nwc{\eea} {\ea\ee}
\nwc{\ben} {\begin{eqnarray}}
\nwc{\een} {\end{eqnarray}}
\nwc{\bda} {\bdm\ba{lcl}}
\nwc{\eda} {\ea\edm}
\nwc{\bc}  {\begin{center}}
\nwc{\ec}  {\end{center}}
\nwc{\ds}  {\displaystyle}
\nwc{\bmat}{\left(\ba}
\nwc{\emat}{\ea\right)}
\nwc{\non} {\nonumber}
\nwc{\bib} {\bibitem}
\nwc{\lra} {\longrightarrow}
\nwc{\Llra}{\Longleftrightarrow}
\nwc{\ra}  {\rightarrow}
\nwc{\Ra}  {\Rightarrow}
\nwc{\lmt} {\longmapsto}
\nwc{\prl} {\partial}
\nwc{\iy}  {\infty}
\nwc{\ol}  {\overline}
\nwc{\hm}  {\hspace{3mm}}
\nwc{\lf}  {\left}
\nwc{\ri}  {\right}
\nwc{\lm}  {\limits}
\nwc{\lb}  {\lbrack}
\nwc{\rb}  {\rbrack}
\nwc{\ov}  {\over}
\nwc{\pr}  {\prime}
\nwc{\nnn} {\nonumber \vspace{.2cm} \\ }
\nwc{\Sc}  {{\cal S}}
\nwc{\Lc}  {{\cal L}}
\nwc{\Rc}  {{\cal R}}
\nwc{\Dc}  {{\cal D}}
\nwc{\Oc}  {{\cal O}}
\nwc{\Cc}  {{\cal C}}
\nwc{\Pc}  {{\cal P}}
\nwc{\Mc}  {{\cal M}}
\nwc{\Ec}  {{\cal E}}
\nwc{\Fc}  {{\cal F}}
\nwc{\Hc}  {{\cal H}}
\nwc{\Kc}  {{\cal K}}
\nwc{\Xc}  {{\cal X}}
\nwc{\Gc}  {{\cal G}}
\nwc{\Zc}  {{\cal Z}}
\nwc{\Nc}  {{\cal N}}
\nwc{\fca} {{\cal f}}
\nwc{\xc}  {{\cal x}}
\nwc{\Ac}  {{\cal A}}
\nwc{\Bc}  {{\cal B}}
\nwc{\Uc}  {{\cal U}}
\nwc{\Vc}  {{\cal V}}
\nwc{\Th} {\Theta}
\nwc{\th} {\theta}
\nwc{\vth} {\vartheta}
\nwc{\eps}{\epsilon}
\nwc{\si} {\sigma}
\nwc{\Gm} {\Gamma}
\nwc{\gm} {\gamma}
\nwc{\bt} {\beta}
\nwc{\La} {\Lambda}
\nwc{\la} {\lambda}
\nwc{\om} {\omega}
\nwc{\Om} {\Omega}
\nwc{\dt} {\delta}
\nwc{\Si} {\Sigma}
\nwc{\Dt} {\Delta}
\nwc{\al} {\alpha}
\nwc{\vph}{\varphi}
\def\tr{\mathop{\rm tr}}
\def\Tr{\mathop{\rm Tr}}

\def\VEV#1{\left\langle #1\right\rangle}

\def\pr#1{#1^\prime}

\nwc{\Id}  {{\bf 1}}
\nwc{\diag} {{\rm diag}}
\nwc{\inv}  {{\rm inv}}
\nwc{\mod}  {{\rm mod}}
\nwc{\hal} {\frac{1}{2}}
\nwc{\tpi}  {2\pi i}

\def\slash#1{#1\!\!\!/\!\,\,}

\def\pr#1{Phys. Rev. {\bf #1}}

\def\MeV {\,{\rm  MeV}}
\def\GeV {\,{\rm  GeV}}

\def \lta {\mathrel{\vcenter
     {\hbox{$<$}\nointerlineskip\hbox{$\sim$}}}}
\def \gta {\mathrel{\vcenter
     {\hbox{$>$}\nointerlineskip\hbox{$\sim$}}}} 
\newsavebox{\nnin} \sbox{\nnin}{$\hspace{1mm}\in\kern -.8em /
                   \hspace{1mm}$}

\newcommand{\sub}{\subset}
\newsavebox{\nnsub} \sbox{\nnsub}{$\hspace{1mm}\sub\kern -.9em /
            \hspace{1mm}$}

\def\KK{{\rm I\kern -.2em  K}}
\def\NN{{\rm I\kern -.16em N}}
\def\RR{{\rm I\kern -.2em  R}}
\def\ZZ{Z \kern -.43em Z}
\def\QQ{{\rm \kern .25em
             \vrule height1.4ex depth-.12ex width.06em\kern-.31em Q}}
\def\CC{{\rm \kern .25em
             \vrule height1.4ex depth-.12ex width.06em\kern-.31em C}}
\def\ZZZ{Z\kern -0.31em Z}

\nwc{\olnu}  {\ol{\nu}}
\nwc{\olla}  {\ol{\la}}
\nwc{\olm}   {\ol{m}}
\nwc{\olq}   {\ol{q}}
\nwc{\olmu}  {\ol{\mu}}
\nwc{\olh}   {\ol{h}}
\nwc{\olpsi} {\ol{\psi}}
\nwc{\olsi}  {\ol{\sigma}}
\nwc{\olgm}  {\ol{\gm}}
\nwc{\prlt}  {\frac{\prl}{\prl t}}
\nwc{\ttau}  {\tilde{\tau}}
\nwc{\tP}    {\tilde{P}}
\nwc{\tU}    {\tilde{U}}
\nwc{\teps}  {\tilde{\eps}}
\nwc{\tla}   {\tilde{\la}}
\nwc{\tit}    {\tilde{t}}
\nwc{\tchi}  {\tilde{\chi}}
\nwc{\iddq}  {\int\frac{d^dq}{(2\pi)^d}}
\nwc{\iddp}  {\int\frac{d^dp}{(2\pi)^d}}
\nwc{\iddQ}  {\int\frac{d^dQ}{(2\pi)^d}}
\nwc{\prpr}  {\prime\prime}
\nwc{\rN}    {\left(\frac{\rho}{N}\right)}
\nwc{\rNt}    {\left(\frac{\rho}{N}\right)^{\frac{N-2}{2}}}
\nwc{\rnN}   {\left(\frac{\rho_0}{N}\right)}
\nwc{\rnNt}    {\left(\frac{\rho_0}{N}\right)^{\frac{N-2}{2}}}
\nwc{\rnNf}    {\left(\frac{\rho_0}{N}\right)^{\frac{N-4}{2}}}
\nwc{\rNs}    {\left(\frac{\rho_0}{N}\right)^{\frac{N-6}{2}}}
\nwc{\kNt}    {\left(\frac{\kappa}{N}\right)^{\frac{N-2}{2}}}
\nwc{\kNf}    {\left(\frac{\kappa}{N}\right)^{\frac{N-4}{2}}}
\nwc{\kNs}    {\left(\frac{\kappa}{N}\right)^{\frac{N-6}{2}}}

\nwc{\cst}     {SU_L(3)\times SU_R(3)}
\nwc{\csN}     {SU_L(N)\times SU_R(N)}
\nwc{\rmcl}    {{\rm cl}}
\nwc{{\rmeff}}   {{\rm eff}}

\begin{document}


\title{Nonperturbative Flow Equations\\ and Low--Energy
  QCD~\footnote{Based on invited talks by the authors at the {\em
      Workshop on Quantum Chromodynamics: Confinement, Collisions, and
      Chaos}, Paris, France, June 1996, and {\em QCD 96}, Yaroslavl,
    Russia, May 1996.}}

\author{{\sc D.--U.~Jungnickel\thanks{Email: 
D.Jungnickel@thphys.uni-heidelberg.de}} \\
 \\ and \\ \\
{\sc C.~Wetterich\thanks{Email: C.Wetterich@thphys.uni-heidelberg.de}} 
\\ \\ \\
{\em Institut f\"ur Theoretische Physik} \\
{\em Universit\"at Heidelberg} \\
{\em Philosophenweg 16} \\
{\em 69120 Heidelberg, Germany}}

\date{}




\thispagestyle{empty}

\maketitle

\begin{picture}(5,2.5)(-350,-500)
\put(12,-115){HD--THEP--96--40}
\put(12,-138){September, 1996}
\end{picture}

\begin{abstract}
  We review the formalism of the effective average action in quantum
  field theory which corresponds to a coarse grained free energy in
  statistical mechanics. The associated exact renormalization group
  equation and possible nonperturbative approximations for its
  solution are discussed. This is applied to QCD where one observes
  the consecutive emergence of mesonic bound states and spontaneous
  chiral symmetry breaking as the coarse graining scale is lowered. We
  finally present a study of the phenomenological importance of
  non--renormalizable terms in the effective linear meson model.
\end{abstract}

\section{Effective average action}
\label{EffectiveAverageAction}

Quantum chromodynamics (QCD) describes qualitatively different physics
at different length scales. At short distances the relevant degrees of
freedom are quarks and gluons which can be treated perturbatively. At
long distances we observe hadrons and an essential part of the
dynamics can be encoded in the masses and interactions of mesons. Any
attempt to deal with this situation analytically and to predict the
meson properties from the short distance physics (as functions of the
strong gauge coupling $\alpha_s$ and the current quark masses $m_q$)
has to bridge the gap between two qualitatively different effective
descriptions. Two basic problems have to be mastered for an
extrapolation from short distance QCD to mesonic length scales:
\begin{itemize}
\item The effective couplings change with scale. This does not only
  concern the running gauge coupling, but also the coefficients of
  non--renormalizable operators as, for example, four quark operators.
  Typically, these non--renormalizable terms become important in the
  momentum range where $\alpha_s$ is strong and deviate substantially
  from their perturbative values. Consider the four--point function
  which obtains after integrating out the gluons. For heavy quarks it
  contains the information about the shape of the heavy quark
  potential whereas for light quarks the complicated spectrum of light
  mesons and chiral symmetry breaking are encoded in it. At distance
  scales around $1{\rm fm}$ one expects that the effective action
  resembles very little the form of the classical QCD action which is
  relevant at short distances.
\item Not only the couplings, but even the relevant variables or
  degrees of freedom are different for long distance and short
  distance QCD. It seems forbiddingly difficult to describe the
  low--energy scattering of two mesons in a language of quarks and
  gluons only. An appropriate analytical field theoretical method
  should be capable of introducing field variables for composite
  objects such as mesons.
\end{itemize}
A conceptually very appealing idea for our task is the block--spin
action~\cite{Kad66-1,Wil71-1}. It realizes that physics with a given
characteristic length scale $l$ is conveniently described by a
functional integral with an ultraviolet (UV) cutoff $\Lambda$ for the
momenta. Here $\Lambda$ should be larger than $l^{-1}$ but not
necessarily by a large factor. The Wilsonian effective action
$S_\Lambda^{\rm W}$ replaces then the classical action in the
functional integral. It obtains by integrating out the fluctuations
with momenta $q^2\gta\Lambda^2$. An exact renormalization group
equation~\cite{Wil71-1}--\cite{Has86-1} describes how
$S_\Lambda^{\rm W}$ changes with the UV cutoff $\Lambda$.

We will use here the somewhat different but related concept of the
effective average action~\cite{Wet91-1} $\Gamma_k$ which, in the
language of statistical physics, is a coarse grained free energy with
coarse graining scale $k$. The effective average action is based on
the quantum field theoretical concept of the effective
action~\cite{Sch51-1} $\Gamma$ which obtains by integrating out all
quantum fluctuations.  The effective action contains all information
about masses, couplings, form factors and so on, since it is the
generating functional of the $1PI$ Green functions. The field
equations derived from $\Gamma$ are exact including all quantum
effects. For a field theoretical description of thermal equilibrium
this concept is easily generalized to a temperature dependent
effective action which includes now also the thermal fluctuations. In
statistical physics $\Gamma$ describes the free energy as a functional
of some convenient (space dependent) order parameter, for instance the
magnetization. In particular, the behavior of $\Gamma$ for a constant
order parameter (the effective potential) specifies the equation of
state. The effective average action $\Gamma_k$ is a simple
generalization of the effective action, with the distinction that only
quantum fluctuations with momenta $q^2\gta\ k^2$ are included. This
can be achieved by introducing in the functional integral defining the
partition function (or the generating functional for the $n$--point
functions) an explicit infrared cutoff $\sim k$.  Typically, this
IR--cutoff is quadratic in the fields and modifies the inverse
propagator, for example by adding a mass--like term $\sim k^2$. The
effective average action can then be defined in complete analogy to
the effective action (via a Legendre transformation of the logarithm of
the partition function). The mass--like term in the propagator
suppresses the contributions from the small momentum modes with
$q^2\lta k^2$ and $\Gamma_k$ accounts effectively only for the
fluctuations with $q^2\gta k^2$.

Following the behavior of $\Gamma_k$ for different $k$ is like looking
at the world through a microscope with variable resolution: For large
$k$ one has a very precise resolution $\sim k^{-1}$ but one also
studies effectively only a small volume $\sim k^d$. Taking in QCD the
coarse graining scale $k$ much larger than the confinement scale
guarantees that the complicated nonperturbative physics does not play
a role yet.  In this case, $\Gamma_k$ will look similar to the
classical action, typically with a running gauge coupling evaluated at
the scale $k$.  (This does not hold for Green functions with much
larger momenta $p^2\gg k^2$ where the relevant IR cutoff is $p$ and
the effective coupling $\alpha_s(p)$.)  For lower $k$ the resolution
is smeared out and the detailed information of the short distance
physics can be lost.  (Again, this does not concern Green functions at
high momenta.)  On the other hand, the ``observable volume'' is
increased and long distance aspects such as collective phenomena
become now visible. In a theory with a physical UV cutoff $\Lambda$ we
may associate $\Gamma_\Lambda$ with the classical action $S$ since no
fluctuations are effectively included. By definition, the effective
average action equals the effective action for $k=0$,
$\Gamma_{0}=\Gamma$, since the infrared cutoff is absent. Thus
$\Gamma_k$ interpolates between the classical action $S$ and the
effective action $\Gamma$ as $k$ is lowered from $\Lambda$ to zero.
The ability to follow the evolution to $k\ra0$ is equivalent to the
ability to solve the quantum field theory.

For a formal description we will consider in the first two sections a
model with real scalar fields $\chi^a$ (the index $a$ labeling
internal degrees of freedom) in $d$ Euclidean dimensions with
classical action $S[\chi]$. We define the generating functional for
the connected Green functions by
\begin{equation}
  W_k[J]=\ln\int\Dc\chi\exp\left\{-S_k[\chi]+\int d^d x
  J_a(x)\chi^a(x)\right\}
\end{equation} 
where we have added to the classical action an IR cutoff $\Delta_kS$
\begin{equation}
  S_k[\chi] = S[\chi]+\Dt S_k[\chi]
\end{equation}
which is quadratic in the fields and best formulated in momentum space 
\begin{equation}
  \Dt S_k[\chi] = \hal\iddq
  R_k(q^2)\chi_a(-q)\chi^a(q)\; .  
\end{equation}
Here $J_a$ are the usual scalar sources introduced to define
generating functionals and $R_k(q^2)$ denotes an appropriately chosen
(see below) IR cutoff function. We require that $R_k(q^2)$ vanishes
rapidly for $q^2\gg k^2$ whereas for $q^2\ll k^2$ it behaves as
$R_k(q^2)\simeq k^2$. This means that all Fourier components of
$\chi^a$ with momenta smaller than the IR cutoff $k$ acquire an
effective mass $m_\rmeff\simeq k$ and therefore decouple while the
high momentum components of $\chi^a$ should not be affected by $R_k$.
The classical fields
\begin{equation}
  \Phi^a\equiv\VEV{\chi^a}=\frac{\dt W_k[J]}{\dt J_a} 
\end{equation} 
depend now on $k$. In terms of $W_k$ the effective average action is
defined via a Legendre transform
\begin{equation}
  \label{AAA60}
  \Gm_k[\Phi]=-W_k[J]+\int d^dx J_a(x)\Phi^a(x)- 
  \Delta S_k[\Phi] \; .
\end{equation}
In order to define a reasonable coarse grained free energy we have
subtracted in (\ref{AAA60}) the infrared cutoff piece. This guarantees
that the only difference between $\Gamma_k$ and $\Gamma$ is the
effective IR cutoff in the fluctuations. Furthermore, this has the
consequence that $\Gamma_k$ does not need to be convex whereas a pure
Legendre transform is always convex by definition. (The coarse grained
free energy becomes convex~\cite{RW90-1} only for $k\ra0$.) This is
very important for the description of phase transitions, in particular
first order ones. One notes
\begin{equation}\begin{array}{lcl}
 \ds{\lim_{k\ra0}R_k(q^2)=0} &\Ra& 
 \ds{\lim_{k\ra0}\Gm_k[\Phi]=\Gm[\Phi]}\nnn
 \ds{\lim_{k\ra\La}R_k(q^2)=\infty} &\Ra&
 \ds{\lim_{k\ra\La}\Gm_k[\Phi]=S[\Phi]}
 \label{ConditionsForRk}
\end{array}\end{equation}
for a convenient choice of $R_k$ like
\begin{equation}
 R_k(q^2)=Z_k q^2
 \frac{1+e^{-q^2/k^2}-e^{-q^2/\La^2}}
 {e^{-q^2/\La^2}-e^{-q^2/k^2}}\; .
 \label{Rk}
\end{equation}
Here $Z_k$ denotes the wave function renormalization to be specified
below and we will often use for $R_k$ the limit $\Lambda\ra\infty$
\begin{equation}
  \label{AAA61}
  R_k(q^2)=\frac{Z_k q^2}
  {e^{q^2/k^2}-1}\; .
\end{equation}
We note that the property $\Gamma_\Lambda=S$ is not essential since
the short distance laws may be parameterized by $\Gamma_\Lambda$ as
well as by $S$. In addition, for momentum scales much smaller than
$\Lambda$ universality implies that the precise form of
$\Gamma_\Lambda$ is irrelevant, up to the values of a few relevant
renormalized couplings.

A few properties of the effective average action are worth mentioning:
\begin{enumerate}
\item All symmetries of the model that are respected by the IR cutoff
  $\Delta_k S$ are automatically symmetries of $\Gamma_k$. In
  particular, this concerns translation and rotation invariance and
  one is not plagued by many of the problems encountered by a
  formulation of the block--spin action on a lattice.
\item In consequence, $\Gamma_k$ can be expanded in terms of
  invariants with respect to these symmetries with couplings depending
  on $k$. For the example of a scalar theory one may use a derivative
  expansion ($\rho=\Phi^a\Phi_a/2$)
  \begin{equation}
    \label{AAA63}
    \Gamma_k=\int d^d x\left\{
    U_k(\rho)+\frac{1}{2}Z_k(\rho)
    \prl^\mu\Phi_a\prl_\mu\Phi^a+\ldots\right\}
  \end{equation}
  and expand further in powers of $\rho$
  \begin{eqnarray}
    \label{AAA64}
    \ds{U_k(\rho)} &=& \ds{
    \frac{1}{2}\ol{\lambda}_k\left(
    \rho-\rho_0(k)\right)^2+
    \frac{1}{6}\ol{\gamma}_k\left(
    \rho-\rho_0(k)\right)^3+\ldots}\nnn
    \ds{Z_k(\rho)} &=& \ds{
    Z_k(\rho_0(k))+Z_k^\prime(\rho_0(k))
    \left(\rho-\rho_0(k)\right)+\ldots}\; .
  \end{eqnarray}
  We see that $\Gamma_k$ describes infinitely many running couplings.
  ($Z_k$ in (\ref{Rk}) can be identified with $Z_k(\rho_0)$.)
\item There is no problem incorporating chiral fermions since a
  chirally invariant cutoff $R_k$ can be formulated~\cite{Wet90-1}.
\item Gauge theories can be formulated along similar
  lines~\cite{RW93-1}--\cite{EHW96-1} even though $\Delta_k S$ may
  not be gauge invariant. In this case the usual Ward identities
  receive corrections for which one can derive closed
  expressions~\cite{EHW94-1}. These corrections vanish for $k\ra0$.
\item The high momentum modes are very effectively integrated out
  because of the exponential decay of $R_k$ for $q^2\gg k^2$.
  Nevertheless, it is sometimes technically easier to use a cutoff
  without this fast decay property (e.g.~$R_k\sim k^2$ or $R_k\sim
  k^4/q^2$). In the latter case one has to be careful with possible
  remnants of an incomplete integration of the short distance modes.
  Also our cutoff does not introduce any non--analytical behavior as
  would be the case for a sharp cutoff~\cite{Wet91-1}.
\item Despite a similar spirit and many analogies there remains also a
  conceptual difference to the Wilsonian effective action
  $S_\Lambda^{\rm W}$. The Wilsonian effective action describes a set
  of different actions (parameterized by $\Lambda$) for one and the
  same theory --- the $n$--point functions are independent of
  $\Lambda$ and have to be computed from $S_\Lambda^{\rm W}$ by
  further functional integration. In contrast, $\Gamma_k$ describes
  the effective action for different theories --- for any value of $k$
  the effective average action is related to the generating functional
  of a theory with a different action $S_k=S+\Delta_k S$. The
  $n$--point functions depend on $k$. The Wilsonian effective action
  does not generate the $1PI$ Green functions~\cite{KKS92-1}.
\item Because of the incorporation of an infrared cutoff, $\Gamma_k$
  is closely related to an effective action for averages of
  fields~\cite{Wet91-1}, where the average is taken over a volume
  $\sim k^d$.
\end{enumerate}

\section{Exact renormalization group equation}
\label{AnExactRGE}

The dependence of $\Gamma_k$ on the coarse graining scale $k$ is
governed by an exact renormalization group equation
(ERGE)~\cite{Wet93-1}
\begin{equation}
  \prl_t\Gm_k[\Phi] =  \hal\Tr\left\{\left[
  \Gm_k^{(2)}[\Phi]+R_k\right]^{-1}\prl_t R_k\right\} \; .
  \label{ERGE}
\end{equation}
Here $t=\ln(k/\La)$ with some arbitrary momentum scale $\La$, and the
trace includes a momentum integration as well as a summation over
internal indices, $\Tr=\int\frac{d^d q}{(2\pi)^d}\sum_a$. The second
functional derivative $\Gm_k^{(2)}$ denotes the {\em exact} inverse
propagator
\begin{equation}
 \label{AAA69}
 \left[\Gm_k^{(2)}\right]_{ab}(q,q^\prime)=
 \frac{\dt^2\Gm_k}{\dt\Phi^a(-q)\dt\Phi^b(q^\prime)}\; .
\end{equation}
The flow equation (\ref{ERGE}) can be derived from (\ref{AAA60}) in a
straightforward way using
\begin{eqnarray}
  \label{AAA70}
  \ds{\prl_t \left.\Gamma_k\right|_\Phi} &=& \ds{
  -\prl_t \left.W_k\right|_J-
  \prl_t\Delta_kS[\Phi]}\nnn
  &=& \ds{
  \frac{1}{2}\Tr\left\{\prl_t R_k\left(
  \VEV{\Phi\Phi}-\VEV{\Phi}\VEV{\Phi}\right)
  \right\}}\nnn
  &=& \ds{
  \frac{1}{2}\Tr\left\{
  \prl_t R_k W^{(2)}_k\right\} }
\end{eqnarray}
and
\begin{eqnarray}
  \label{AAA71}
  \ds{W_{k,ab}^{(2)}(q,q^\prime)} &=& \ds{
  \frac{\delta^2 W_k}
  {\delta J^a(-q)\delta J^b(q^\prime)} }\nnn
  \ds{\frac{\delta^2 W_k}
  {\delta J_a(-q)\delta J_b(q^\prime)}
  \frac{\delta^2\left(\Gamma_k+\Delta_k S\right)}
  {\delta\Phi_b(-q^{\prime})\delta\Phi_c
  (q^{\prime\prime})}  } &=& \ds{
  \delta_{ac}\delta_{q q^{\prime\prime}} }\; .
\end{eqnarray}
It has the form of a renormalization group improved one--loop
expression~\cite{Wet91-1}. Indeed, the one--loop formula for
$\Gamma_k$ reads
\begin{equation}
  \label{AAA72}
  \Gamma_k[\Phi]=S[\Phi]+
  \frac{1}{2}\Tr\ln\left(
  S^{(2)}[\Phi]+R_k\right)
\end{equation}
with $S^{(2)}$ the second functional derivative of the classical
action, similar to (\ref{AAA69}).  (Remember that $S^{(2)}$ is the
field dependent classical inverse propagator. Its first and second
derivative with respect to the fields describe the classical three--
and four--point vertices, respectively.) Taking a $t$--derivative of
eq.~(\ref{AAA72}) gives a one--loop flow equation very similar to
(\ref{ERGE}) with $\Gamma_k^{(2)}$ replaced by $S^{(2)}$. It may seem
surprising, but it is nevertheless true, that the renormalization
group improvement $S^{(2)}\ra\Gamma_k^{(2)}$ promotes the one--loop
flow equation to an exact nonperturbative flow equation which includes
the effects from all loops as well as all contributions which are
non--analytical in the couplings like instantons, etc.! For practical
computations it is actually often quite convenient to write the flow
equation (\ref{ERGE}) as a formal derivative of a renormalization
group improved one--loop expression
\begin{equation}
  \label{AAA73}
  \prl_t\Gamma_k=
  \frac{1}{2}\Tr\tilde{\prl}_t\ln\left(
  \Gamma_k^{(2)}+R_k\right)
\end{equation}
with $\tilde{\prl}_t$ acting only on $R_k$ and not on
$\Gamma_k^{(2)}$, i.e. $\tilde{\prl}_t=\left(\prl R_k/\prl
t\right)\left(\prl/\prl R_k\right)$. Flow equations for $n$--point
functions follow from appropriate functional derivatives of
(\ref{ERGE}) or (\ref{AAA73}) with respect to the fields. For their
derivation it is sufficient to evaluate the corresponding one--loop
expressions (with the vertices and propagators derived from
$\Gamma_k$) and then to take a formal $\tilde{\prl}_t$--derivative.
(If the one--loop expression is finite or properly regularized the
$\tilde{\prl}_t$--derivative can be taken after the evaluation of the
trace.) This permits the use of (one--loop) Feynman diagrams and
standard perturbative techniques in many circumstances. Most
importantly, it establishes a very direct connection between the
solution of flow--equations and perturbation theory. If one uses on
the right hand side of eq.~(\ref{ERGE}) a truncation for which the
propagator and vertices appearing in $\Gamma_k^{(2)}$ are replaced by
the ones derived from the classical action, but with running
$k$--dependent couplings, and then expands the result to lowest
non--trivial order in the coupling constants one recovers standard
renormalization group improved one--loop perturbation theory. The
formal solution of the flow equation can also be employed for the
development of a systematically resummed perturbation
theory~\cite{Wet96-1}.

For a choice of the cutoff function similar to (\ref{AAA61}) the
momentum integral contained in the trace on the right hand side of the
flow equation is both infrared and ultraviolet finite. Infrared
finiteness arises through the presence of the infrared regulator $\sim
R_k$. We note that all eigenvalues of the matrix $\Gamma_k^{(2)}+R_k$
must be positive semi--definite. The proof follows from the
observation that the functional $\Gamma_k+\Delta_k S$ is convex since
it is obtained from $W_k$ by a Legendre transform. On the other hand,
ultraviolet finiteness is related to the fast decay of $\prl_t R_k$
for $q^2\gg k^2$. This expresses the fact that only a narrow range of
fluctuations with $q^2\simeq k^2$ contributes effectively if the
infrared cutoff $k$ is lowered by a small amount~\footnote{If for some
  other choice of $R_k$ the right hand side of the flow equation would
  not remain UV finite this would indicate that the high momentum
  modes have not yet been integrated out completely in the computation
  of $\Gamma_k$.}.  Since the flow equation is manifestly finite this
can be used to define a regularization scheme. The ``ERGE--scheme'' is
specified by the flow equation, the choice of $R_k$ and the ``initial
condition'' $\Gamma_\Lambda$. This is particularly important for gauge
theories where other regularizations in four dimensions and in the
presence of chiral fermions are difficult to construct~\footnote{For
  gauge theories $\Gamma_\Lambda$ has to obey appropriately modified
  Ward identities. In the context of perturbation theory a first
  proposal how to regularize gauge theories by use of flow equations
  can be found in~\cite{Bec96-1,BAM94-1}.}. We note that in contrast
to previous versions of exact renormalization group equations there is
no need in the present formulation to construct an ultraviolet
momentum cutoff --- a task known to be very difficult in non--Abelian
gauge theories.

Despite the conceptual differences between the Wilsonian effective
action $S_\Lambda^{\rm W}$ and the effective average action
$\Gamma_k$ the exact flow equations describing the $\Lambda$--dependence
of $S_\Lambda^{\rm W}$ and the $k$--dependence of $\Gamma_k$ are
simply related. Polchinski's continuum version of the Wilsonian flow
equation~\cite{Pol84-1} can be transformed into (\ref{ERGE}) by means
of a Legendre transform and a suitable variable
redefinition~\cite{BAM93-1}.

Even though intuitively simple, the replacement of the (RG--improved)
classical propagator by the full propagator turns the solution of the
flow equation (\ref{ERGE}) into a difficult mathematical problem: The
evolution equation is a functional differential equation. Once
$\Gamma_k$ is expanded in terms of invariants (e.g.~(\ref{AAA63}),
(\ref{AAA64})) this is equivalent to a coupled system of non--linear
partial differential equations for infinitely many couplings. General
methods for the solution of functional differential equations are not
developed very far. They are mainly restricted to iterative procedures
that can be applied once some small expansion parameter is identified.
This covers usual perturbation theory in the case of a small coupling,
the $1/N$--expansion or expansions in the dimensionality $4-d$ or
$2-d$. It may also be extended to less familiar expansions like a
derivative expansion which is related in critical three dimensional
scalar theories to a small anomalous dimension $\eta$. In the absence
of a clearly identified small parameter one nevertheless needs to
truncate the most general form of $\Gamma_k$ in order to reduce the
infinite system of coupled differential equations to a (numerically)
manageable size. This truncation is crucial. It is at this level that
approximations have to be made and, as for all nonperturbative
analytical methods, they are often not easy to control. The challenge
for nonperturbative systems like low momentum QCD is to find flow
equations which (a) incorporate all the relevant dynamics such that
neglected effects make only small changes, and (b) remain of
manageable size. The difficulty with the first task is a reliable
estimate of the error. For the second task the main limitation is a
practical restriction for numerical solutions of differential
equations to functions depending only on a small number of variables.
The existence of an exact functional differential flow equation is a
very useful starting point and guide for this task. At this point the
precise form of the exact flow equation is quite important.
Furthermore, it can be used for systematic expansions through
enlargement of the truncation and for an error estimate in this way.
Nevertheless, this is not all. Usually, physical insight into a model
is necessary to device a useful nonperturbative truncation!

So far, two complementary approaches to nonperturbative truncations
have been explored: an expansion of the effective Lagrangian in powers
of derivatives ($\rho\equiv\frac{1}{2}\Phi_a\Phi^a$)
\begin{equation}
\label{DerExp}
  \Gamma_k[\Phi]=\int d^d x\left\{
  U_k(\rho)+\frac{1}{2}Z_{k}(\Phi)
  \prl_\mu\Phi^a\prl^\mu\Phi_a+
  \frac{1}{4}Y_{k}(\rho)\prl_\mu\rho
  \prl^\mu\rho+
  \Oc(\prl^4)\right\}
\end{equation}
or one in powers of the fields
\begin{equation}
\label{FieldExp}
  \Gamma_k[\Phi]=
  \sum_{n=0}^\infty\frac{1}{n!}\int
  \left(\prod_{j=0}^n d^d x_j
  \left[\Phi(x_j)-\Phi_0\right]\right)
  \Gamma_k^{(n)}(x_1,\ldots,x_n)\; .
\end{equation}
If one chooses $\Phi_0$ as the $k$--dependent VEV of $\Phi$, the
series (\ref{FieldExp}) starts effectively at $n=2$. The flow
equations for the $1PI$ $n$--point functions $\Gamma_k^{(n)}$ are
obtained by functional differentiation of
eq.~(\ref{ERGE})~\footnote{Such flow equations have been discussed
  earlier from a somewhat different viewpoint~\cite{Wei76-1}. They can
  also be interpreted as a differential form of Schwinger--Dyson
  equations~\cite{DS49-1}.}.  The formation of mesonic bound states,
which typically appear as poles in the (Minkowskian) four quark Green
function, is most efficiently described by expansions like
(\ref{FieldExp}).  This is also the form needed to compute the
nonperturbative momentum dependence of the gluon propagator and the
heavy quark potential~\cite{Wet95-2,EHW96-1}.  On the other hand, a
parameterization of $\Gamma_k$ as in (\ref{DerExp}) seems particularly
suited for the study of phase transitions. The evolution equation for
the average potential $U_k$ follows by evaluating (\ref{DerExp}) for
constant $\Phi$. In the limit where the $\Phi$--dependence of
$Z_{k}$ is neglected and $Y_{k}=0$ one finds~\cite{Wet91-1}
for the $O(N)$--symmetric scalar model
\begin{equation}
  \label{AAA80}
  \prl_t U_k(\rho)=\frac{1}{2}
  \int\frac{d^d q}{(2\pi)^d}
  \frac{\prl R_k}{\prl t}\left(
  \frac{N-1}{Z_{k}q^2+R_k+U_k^\prime}+
  \frac{1}{Z_{k}q^2+R_k(q)+U_k^\prime+
  2\rho U_k^{\prime\prime}}\right)
\end{equation}
with $U_k^\prime\equiv\frac{\prl U_k}{\prl\rho}$, etc. One observes
the appearance of $\rho$--dependent mass terms in the effective
propagators of the right hand side of eq.~(\ref{AAA80}). Once
$\eta_k\equiv-\prl_t\ln Z_{k}$ is determined~\cite{Wet91-1} in terms
of the couplings parameterizing $U_k$ this is a partial differential
equation~\footnote{In the sharp cutoff limit and for $\eta_k=0$ the
  flow equation for $U_k$ coincides with the one corresponding to the
  Wilsonian effective action~\cite{Has86-1}. Because of
  non--analyticities in the kinetic terms~\cite{Wet91-1} an analysis
  of the derivative terms is difficult for a sharp cutoff.} for a
function $U_k$ depending on two variables $k$ and $\rho$ which can be
solved numerically~\cite{ABB95-1}--\cite{BW96-1}. (The
Wilson--Fisher fixed point relevant for a second order phase
transition ($d=3$) corresponds to a scaling
solution~\cite{TW94-1,Mor94-1} where $\prl_t U_k=0$.) A suitable
truncation of a flow equation of the type (\ref{DerExp}) will play a
central role in the description of chiral symmetry breaking below.

It should be mentioned at this point that the weakest point in the
ERGE approach seems to be a reliable estimate of the truncation error
in a nonperturbative context. This problem is common to all known
analytical approaches to nonperturbative phenomena and appears often
even within systematic (perturbative) expansions. One may hope that
the existence of an exact flow equation could also be of some help for
error estimates. An obvious possibility to test a given truncation is
its enlargement to include more variables --- for example, going one
step higher in a derivative expansion. This is similar to computing
higher orders in perturbation theory and limited by practical
considerations. As an alternative, one may employ different
truncations of comparable size --- for instance, by using different
definitions of retained couplings. A comparison of the results can
give a reasonable picture of the uncertainty if the used set of
truncations is wide enough. In this context we should also note the
dependence of the results on the choice of the cutoff function
$R_k(q)$.  Of course, for $k\ra0$ the physics should not depend on a
particular choice of $R_k$ and, in fact, it does not for full
solutions of (\ref{ERGE}). Different choices of $R_k$ just correspond
to different trajectories in the space of effective actions along
which the unique IR limit $\Gamma[\Phi]$ is reached. Once
approximations are used to solve the ERGE (\ref{ERGE}), however, not
only the trajectory but also its end point will depend on the precise
definition of the function $R_k$. This is very similar to the
renormalization scheme dependence usually encountered in perturbative
computations of Green functions. One may use this scheme dependence as
a tool to study the robustness of a given approximation scheme.

Before applying a new nonperturbative method to a complicated theory
like QCD it should be tested for simpler models. A good criterion for
the capability of the ERGE to deal with nonperturbative phenomena
concerns the critical behavior in three dimensional scalar theories.
In a first step the well known results of other methods for the
critical exponents have been reproduced within a few percent
accuracy~\cite{TW94-1}. The ability of the method to produce new
results has been demonstrated by the computation of the critical
equation of state for Ising and Heisenberg models~\cite{BTW96-1} which
has been verified by lattice simulations~\cite{Tsy94-1}. This has been
extended to first order transitions in matrix models~\cite{BW96-1} or
for the Abelian Higgs model relevant for
superconductors~\cite{BLL95-1,Tet96-1}.  Analytical investigations of
the high temperature phase transitions in $d=4$ scalar theories
($O(N)$--models) have correctly described the second order nature of
the transition~\cite{TW93-1}, in contrast to earlier attempts within high
temperature perturbation theory.

For an extension of the flow equations to Abelian and non--Abelian
gauge theories we refer the reader
to~\cite{RW93-1}--\cite{EHW96-1}. The other
necessary piece for a description of low--energy QCD, namely the
transition from fundamental (quark and gluon) degrees of freedom to
composite (meson) fields within the framework of the ERGE can be found
in~\cite{EW94-1}. We will describe the most important aspects of this
formalism for mesons below.

\section{Chiral symmetry breaking in QCD}
\label{ChiralSymmetryBreaking}

The strong interaction dynamics of quarks and gluons at short
distances or high energies is successfully described by quantum
chromodynamics (QCD). One of its most striking features is asymptotic
freedom~\cite{GW73-1} which makes perturbative calculations reliable
in the high energy regime. On the other hand, at scales around a few
hundred $\MeV$ confinement sets in. As a consequence, the low--energy
degrees of freedom in strong interaction physics are mesons, baryons
and glueballs rather than quarks and gluons. When constructing
effective models for these IR degrees of freedom one usually relies on
the symmetries of QCD as a guiding principle, since a direct
derivation of such models from QCD is still missing. The most
important symmetry of QCD, its local color $SU(3)$ invariance, is of
not much help here, since the IR spectrum appears to be color neutral.
When dealing with bound states involving heavy quarks the so called
``heavy quark symmetry'' may be invoked to obtain approximate symmetry
relations between IR observables~\cite{Neu94-1}. We will rather focus
here on the light scalar and pseudoscalar meson spectrum and therefore
consider QCD with only the light quark flavors $u$, $d$ and $s$. To a
good approximation the masses of these three flavors can be considered
as small in comparison with other typical strong interaction scales.
One may therefore consider the chiral limit of QCD (vanishing current
quark masses) in which the classical QCD Lagrangian does not couple
left-- and right--handed quarks. It therefore exhibits a global chiral
invariance under $U_L(N)\times U_R(N)=SU_L(N)\times SU_R(N)\times
U_V(1)\times U_A(1)$ where $N$ denotes the number of massless quarks
($N=2$ or $3$) which transform as
\begin{eqnarray}
  \ds{\psi_R\equiv\frac{1-\gm_5}{2}\psi} &\longrightarrow&
  \ds{\Uc_R \psi_R\; ;\;\;\;\Uc_R\in U_R(N)}\nnn
  \ds{\psi_L\equiv\frac{1+\gm_5}{2}\psi} &\longrightarrow& \ds{\Uc_L
  \psi_L\; ;\;\;\;\Uc_L\in U_L(N)}\; .
 \label{ChiralTransoformation}
\end{eqnarray}
Even for vanishing quark masses only the diagonal $SU_V(N)$ vector--like
subgroup can be observed in the hadron spectrum (``eightfold
way''). The symmetry $SU_L(N)\times SU_R(N)$ must therefore be
spontaneously broken to $SU_V(N)$
\begin{equation}
  SU_L(N)\times SU_R(N)\longrightarrow
  SU_{L+R}(N)\equiv SU_V(N)\; .
  \label{CSBPattern}
\end{equation}
Chiral symmetry breaking is one of the most prominent features of
strong interaction dynamics and phenomenologically well
established~\cite{Leu95-1}, even though a rigorous derivation of this
phenomenon starting from first principles is still missing. In
particular, the chiral symmetry breaking (\ref{CSBPattern}) predicts
for $N=3$ the existence of eight light parity--odd (pseudo--)Goldstone
bosons: $\pi^0$, $\pi^\pm$, $K^0$, $\ol{K}^0$, $K^\pm$ and $\eta$.
Their comparably small masses are a consequence of the explicit chiral
symmetry breaking due to small but non--vanishing current quark
masses.  The axial Abelian subgroup $U_A(1)=U_{L-R}(1)$ is broken in
the quantum theory by an anomaly of the axial--vector current. This
breaking proceeds without the occurrence of a Goldstone
boson~\cite{Hoo86-1}.  Finally, the $U_V(1)=U_{L+R}(1)$ subgroup
corresponds to baryon number conservation.

The light pseudoscalar and scalar mesons are thought of as color
neutral quark--antiquark bound states $\Phi^{ab}\sim \ol{\psi}_L^b
\psi_R^a$, $a,b=1,\ldots,N$, which therefore transform under chiral
rotations (\ref{ChiralTransoformation}) as
\begin{equation} 
  \Phi\longrightarrow
  \Uc_R\Phi\Uc_L^\dagger\; .  
\end{equation} 
Hence, the chiral symmetry breaking pattern (\ref{CSBPattern}) is
realized if the meson potential develops a VEV
\begin{equation}
  \label{AAA50}
  \VEV{\Phi^{ab}}=\ol{\sigma}_0\dt^{ab}\;
  ;\;\;\; \ol{\sigma}_0\neq0\; .  
\end{equation} 
One of the most crucial and yet unsolved problems of strong
interaction dynamics is to derive an effective field theory for the
mesonic degrees of freedom directly from QCD which exhibits this
behavior.

\section{A semi--quantitative picture}
\label{ASemiQuantitativePicture}

Before turning to a quantitative description of chiral symmetry
breaking using flow equations it is perhaps helpful to give a brief
overview of the relevant scales which appear in relation to this
phenomenon and the physical degrees of freedom associated to them.
Some of this will be explained in more detail in the remainder of this
talk whereas other parts are rather well established features of
strong interaction physics.

At scales above approximately $1.5\GeV$, the relevant degrees of
freedom of strong interactions are quarks and gluons and their
dynamics appears to be well described by perturbative QCD. At somewhat
lower energies this changes dramatically. Quark and gluon bound states
form and confinement sets in. Concentrating on the physics of scalar
and pseudoscalar mesons~\footnote{One may assume that all other bound
  states are integrated out. We will comment on this issue below.}
there are three important momentum scales which appear to be rather
well separated:
\begin{itemize}
\item The compositeness scale $k_\Phi$ at which mesonic
  $\ol{\psi}\psi$ bound states form due to the increasing strength of
  the strong interaction. It will turn out to be somewhere in the
  range $(600-700)\MeV$.
\item The chiral symmetry breaking scale $k_{\chi SB}$ at which the
  chiral condensate $\VEV{\ol{\psi}^b\psi^a}$ or $\VEV{\Phi^{ab}}$
  will assume a non--vanishing value, therefore breaking chiral
  symmetry according to (\ref{CSBPattern}).  This scale is
  found to be around $(400-500)\MeV$. Below it the quarks acquire
  constituent masses $M_q\simeq350\MeV$ due to their Yukawa coupling
  to the chiral condensate (\ref{AAA50}).
\item The confinement scale $\Lambda_{\rm QCD}\simeq200\MeV$ which
  corresponds to the Landau pole in the perturbative evolution of the
  strong coupling constant $\alpha_s$. In our context, this is the
  scale where deviations of the effective quark propagator from its
  classical form and multi--quark interactions not included in the
  meson physics become of crucial importance.
\end{itemize}
For scales $k$ in the range $k_{\chi SB}\lta k\lta k_\Phi$ the most
relevant degrees of freedom are mesons and quarks. Typically, the
dynamics in this range is dominated by the strong Yukawa coupling $h$
between quarks and mesons: $h^2(k)/(4\pi)\gg\alpha_s(k)$.  One may
therefore assume that the dominant QCD effects are included in the
meson physics and consider a simple model of quarks and mesons only.
As one evolves to scales below $k_{\chi SB}$ the Yukawa coupling
decreases whereas $\alpha_s$ increases. Of course, getting closer to
$\Lambda_{\rm QCD}$ it is no longer justified to neglect the QCD
effects which go beyond the dynamics of effective meson degrees of
freedom~\footnote{A formalism for the inclusion of ``residual'' gluon
  contributions to the flow equation in a situation where part of the
  gluon effects are already described by mesonic bound states can be
  found in~\cite{Wet95-2}.}.  On the other hand, the final IR value of
the Yukawa coupling $h$ is fixed by the typical values of constituent
quark masses $M_q\simeq350\MeV$ to be $h^2/(4\pi)\simeq4.5$. One may
therefore speculate that the domination of the Yukawa interaction
persists down to scales $k\simeq M_q$ at which the quarks decouple
from the evolution of the mesonic degrees of freedom altogether due to
their mass. Of course, details of the gluonic interactions are
expected to be crucial for an understanding of quark and gluon
confinement. Strong interaction effects may dramatically change the
momentum dependence of the quark and gluon propagators for $k$ around
$\Lambda_{\rm QCD}$.  Yet, as long as one is only interested in the
dynamics of the mesons one is led to expect that these effects are
quantitatively no too important. Because of the effective decoupling
of the quarks and therefore the whole colored sector the details of
confinement have only little influence on the mesonic flow equations
for $k\lta\Lambda_{\rm QCD}$.

Let us imagine that we integrate out the gluon degrees of freedom
while keeping an effective infrared cutoff $k_p\simeq1.5\GeV$ in the
quark propagators. The exact flow equation to be used for this purpose
obtains by keeping for $k<k_p$ the infrared cutoff $R$ for the quarks
fixed while lowering the one for the gluons to zero~\footnote{One may
  also only partly integrate out the gluons by stopping this evolution
  at a scale somewhat above $\Lambda_{\rm QCD}$. The remaining
  residual gluon fluctuation can then be included later
  following~\cite{Wet95-2}.}. Subsequently, the gluons are eliminated
by solving the field equations for the gluon fields as functionals of
the quarks. This will result in a non--trivial momentum dependence of
the quark propagator and effective non--local four and higher quark
interactions. Because of the infrared cutoff the resulting effective
action for the quarks resembles closely the one for heavy quarks (at
least for Euclidean momenta). The dominant effect is the appearance of
an effective quark potential (similar to the one for the charm quark)
which describes the effective four quark interactions.  For the
effective quark action at $k_p$ we only retain this four quark
interaction in addition to the classical two--point function, while
neglecting $n$--point functions involving six and more quarks. Details
of the quark two--point function for $q^2\lta\Lambda_{\rm QCD}^2$ will
not be important as long as $k>\Lambda_{\rm QCD}$, and similarly for
the four--point function. For $k<\Lambda_{\rm QCD}$ the quarks
decouple from the mesonic sector as discussed above.  For typical
momenta larger than $\Lambda_{\rm QCD}$ a reliable computation of the
effective quark action should be possible by using in the quark--gluon
flow equation relatively simple truncations~\cite{Wet95-2,EHW96-1}.

We next have to remove the infrared cutoff for the quarks, $k\ra0$.
This task can be carried out by means of the exact flow equation for
quarks only, starting at $k_p$ with an initial value
$\Gamma_{k_p}[\psi]$ as obtained after integrating out the gluons. For
fermions the trace in eq.~(\ref{ERGE}) has to be replaced by a
supertrace in order to account for the minus sign related to Grassmann
variables~\cite{Wet90-1}.  A first investigation in this
direction~\cite{EW94-1} has used a truncation with a chirally
invariant four quark interaction whose most general momentum
dependence was retained
\begin{eqnarray}
 \ds{\Gm_k} &=& \ds{\int\frac{d^4 p}{(2\pi)^4}
 \ol{\psi}_a^i(p)Z_{\psi,k}(p)\left[
 \slash{p}\delta^{ab}+m^{ab}(p)\gamma_5+
 i\tilde{m}^{ab}(p)\right]\psi_{ib}(p)}\nnn 
  &+& \ds{
 \frac{1}{2}\int\left(\prod_{l=1}^4
 \frac{d^4 p_l}{(2\pi)^4}\right)
 \left(2\pi\right)^4\delta(p_1+p_2-p_3-p_4)}\nnn
 &\times& \ds{
 \lambda_k^{(\psi)}(p_1,p_2,p_3,p_4)
 \left\{
 \left[\ol{\psi}_a^i(-p_1)\psi_i^b(p_2)\right]
 \left[\ol{\psi}_b^j(p_4)\psi_j^a(-p_3)\right]
 \right. }\nnn
 && \ds{\left.\hspace{2.95cm}
 -\left[\ol{\psi}_a^i(-p_1)\gamma_5\psi_i^b(p_2)\right]
 \left[\ol{\psi}_b^j(p_4)\gamma_5\psi_j^a(-p_3)\right]
 \right\} }\; .
 \label{QCDFourFermi}
\end{eqnarray}
Here $i,j$ run from one to $N_c$ which is the number of quark colors.
The indices $a,b$ denote different light quark flavors and run from
$1$ to $N$. The matrices $m$ and $\tilde{m}$ are hermitian and
$m+\tilde{m}\gamma_5$ forms therefore the most general quark mass
matrix~\footnote{Our chiral conventions~\cite{Wet90-1} where the
  hermitean part of the mass matrix is multiplied by $\gamma_5$ may be
  somewhat unusual but they are quite convenient for Euclidean
  calculations.}. The ansatz (\ref{QCDFourFermi}) does not correspond
to the most general chirally invariant four quark interaction. It
neglects similar interactions in the $\rho$--meson and pomeron
channels which are also obtained from a Fierz transformation of the
heavy quark potential~\cite{Wet95-2}. With $V(q^2)$ the heavy quark
potential in a Fourier representation, the initial value at
$k_p=1.5\GeV$ is given by ($\hat{Z}_{\psi,k}=Z_{\psi,k}(p^2=-k_p^2)$)
\begin{equation}
  \label{FFCBC}
  \lambda_{k_p}^{(\psi)}(p_1,p_2,p_3,p_4)
  \hat{Z}_{\psi,k_p}^{-2}=
  \frac{1}{2}V((p_1-p_3)^2)=
  \frac{2\pi\alpha_s}{(p_1-p_3)^2}+
  \frac{8\pi\lambda}{\left((p_1-p_3)^2\right)^2}\; .
\end{equation}
For simplicity, the effective heavy quark potential is approximated
here by a one gluon exchange term $\sim\alpha_s(k_p)$ and a string
tension $\lambda\simeq0.18\GeV^2$, but one may use as well a more
precise form of the potential as determined from the phenomenology of
charmonium or as computed from the flow equations with
gluons~\cite{Wet95-2,EHW96-1}. In the simplified ansatz (\ref{FFCBC})
the string tension introduces a second scale in addition to $k_p$ and
it becomes clear that the incorporation of gluon fluctuations is a
crucial ingredient for the emergence of mesonic bound states. For a
more realistic treatment of the heavy quark potential this scale is
set by the running of $\alpha_s$ or $\Lambda_{\rm QCD}$.

The evolution equation for the function $\lambda_k^{(\psi)}$ can be
derived from the fermionic version of eq.~(\ref{ERGE}) and the
truncation (\ref{QCDFourFermi}). Since $\lambda_k^{(\psi)}$ depends on
six independent momentum invariants it is a partial differential
equation for a function depending on seven variables and has to be
solved numerically~\cite{EW94-1}.  The ansatz (\ref{FFCBC})
corresponds to the $t$--channel exchange of a colored gluonic state
and it is by far not clear that the evolution of $\lambda_k^{(\psi)}$
will lead at lower scales to a momentum dependence representing the
exchange of colorless mesonic bound states. Yet, at the compositeness
scale
\begin{equation}
\label{kphi}
  k_\Phi\simeq630\MeV
\end{equation}
one finds an approximate factorization
\begin{equation}
\label{BSFact}
  \lambda_{k_\Phi}^{(\psi)}(p_1,p_2,p_3,p_4)=
  g(p_1,p_2)\tilde{G}(s)g(p_3,p_4)+\ldots
\end{equation}
which indicates the formation of mesonic bound states.  Here
$g(p_1,p_2)$ denotes the amputated Bethe--Salpeter wave function and
$\tilde{G}(s)$ is the mesonic bound state propagator displaying a
pole--like structure in the $s$--channel if it is continued to
negative $s=(p_1+p_2)^2$. The dots indicate the part of
$\lambda_k^{(\psi)}$ which does not factorize and which will be
neglected in the following. In the limit where the momentum dependence
of $g$ and $\tilde{G}$ is neglected we recover the four quark
interaction of the Nambu--Jona-Lasinio model~\cite{NJL61-1,Bij95-1}.
It is therefore not surprising that our description of the dynamics
for $k<k_\Phi$ will parallel certain aspects of the investigations of
this model, even though we are not bound to the approximations used
typically in such studies (large--$N_c$ expansion, perturbative
renormalization group, etc.).

It is clear that for scales $k\lta k_\Phi$ a description of strong
interaction physics in terms of quark fields alone would be rather
inefficient. Finding physically reasonable truncations of the
effective average action should be much easier once composite fields
for the mesons are introduced.  The exact renormalization group
equation can indeed be supplemented by an exact formalism for the
introduction of composite field variables or, more generally, a change
of variables~\cite{EW94-1}. For our purpose, this amounts in practice
to inserting at the scale $k_\Phi$ the identities~\footnote{We use
  here the shorthand notation $A^\dagger G B\equiv\int\frac{d^d
    q}{(2\pi)^d}A_a^*(q)G^{ab}(q) B_b(q)$.}
\begin{eqnarray}
 1 &=& \displaystyle{
 {\rm const}\;
 \int{\cal D}\sigma_A}\nnn
 &\times& \ds{
 \exp\left\{ -\frac{1}{2} \tr
 \left(\sigma_A^\dagger -
 K_A^\dagger \tilde{G}-m_A^\dagger-
 {\cal O}^\dagger \tilde{G}\right)
 \frac{1}{\tilde{G}}
 \left(\sigma_A -\tilde{G}K_A -m_A-
 \tilde{G}{\cal O} \right)\right\} }\nnn
 \label{identity}
  1 &=& \displaystyle{
 {\rm const}\;
 \int{\cal D}\sigma_H}\\[2mm]
 &\times& \displaystyle{
 \exp\left\{-\frac{1}{2}\tr
 \left(\sigma_H^\dagger -
 K_H^\dagger \tilde{G}-m_H^\dagger-
 {\cal O}^{(5)\dagger} \tilde{G}\right)
 \frac{1}{\tilde{G}}\left(\sigma_H -\tilde{G}K_H -m_H-
 \tilde{G}{\cal O}^{(5)} \right)\right\} }\nonumber
\end{eqnarray}
into the functional integral which formally defines the quark
effective average action. Here $K_{A,H}$ are sources for the
collective fields $\sigma_{A,H}$ which correspond in turn to the
anti-hermitian and hermitian parts~\footnote{The fields
  $\overline{\sigma}_{A,H}$ associated to $K_{A,H}$ by a Legendre
  transformation obey $\overline{\sigma}_A^T =-\frac{i}{2}(\Phi
  -\Phi^\dagger )$, $\overline{\sigma}_H^T =\frac{1}{2} (\Phi
  +\Phi^\dagger )$.} of the meson field $\Phi$. They are associated to
the fermion bilinear operators ${\cal O} [\psi]$, ${\cal O}^{(5)}[\psi
]$ whose Fourier components read
\begin{eqnarray}
 \ds{{\cal O}_{\;\; b}^a (q)} &=& \ds{ 
 -i\int\frac{d^4 p}{(2\pi)^4} g(-p,p+q)
 \overline{\psi}^a (p)\psi_b (p+q) }\nnn
 \ds{{\cal O}_{\;\;\;\;\;\; b}^{(5)a} (q)} &=& \ds{ 
 -\int\frac{d^4 p}{(2\pi)^4} g(-p,p+q)
 \overline{\psi}^a (p)\gamma_5\psi_b (p+q) }\; .
\end{eqnarray}
The choice of $g(-p,p+q)$ as the bound state wave function
renormalization and of $\tilde{G}(q)$ as its propagator guarantees
that the four--quark interaction contained in (\ref{identity}) cancels
the dominant factorizing part of the QCD--induced non--local
four--quark interaction (\ref{QCDFourFermi}), (\ref{BSFact}). In
addition, one may choose
\begin{eqnarray}
 \ds{m_{Hab}^T} &=& \ds{
 m_{ab}(0)g^{-1}(0,0)Z_{\psi,k_\Phi}(0)}\nnn
 \ds{m_{Aab}^T} &=& \ds{
 \tilde{m}_{ab}(0)g^{-1}(0,0)Z_{\psi,k_\Phi}(0)}
\end{eqnarray}
such that the explicit quark mass term cancels out for $q=0$. The
remaining quark bilinear, which is proportional to
$m(q)-m(0)Z_{\psi,k_\Phi}(0)g(-q,q)/ [Z_{\psi,k_\Phi}(q)g(0,0)]$ and
breaks chiral symmetry, vanishes for zero momentum and will be
neglected in the following. Without loss of generality we can take $m$
real and diagonal and $\tilde{m}=0$. In consequence, we have replaced
at the scale $k_\Phi$ the effective quark action (\ref{QCDFourFermi})
with (\ref{BSFact}) by an effective quark meson action given by
\begin{eqnarray}
 \ds{\hat{\Gamma}_k} &=& \ds{
 \Gamma_k-\frac{1}{2}\int d^4 x\tr
 \left(\Phi^\dagger\jmath+\jmath^\dagger\Phi\right)}\nnn
 \ds{\Gamma_{k}} &=& \displaystyle{ \int d^4 x
 U_k(\Phi,\Phi^\dagger)
  \label{EffActAnsatz}
 }\\[2mm]
 &+& \displaystyle{ 
 \int\frac{d^4 q}{(2\pi)^d}\Bigg\{
 Z_{\Phi ,k}(q) q^2 \tr\left[
 \Phi^\dagger (q)\Phi (q)\right] +
 Z_{\psi ,k}(q)\overline{\psi}_a(q)
 \gamma^\mu q_\mu \psi^a (q)
 }\nonumber\vspace{.2cm}\\
 &+& \displaystyle{
 \int\frac{d^4 p}{(2\pi)^d}\overline{h} _k (-q,q-p)
 \overline{\psi}^a(q) \left(
 \frac{1+\gamma_5}{2}\Phi _{ab}(p)-
 \frac{1-\gamma_5}{2}\Phi_{ab}^\dagger (-p) \right)
 \psi^b (q-p) \Bigg\}\nonumber \; .}
\end{eqnarray}
At the scale $k_\Phi$ the
inverse scalar propagator is related to $\tilde{G}(q)$ in
eq.~(\ref{BSFact}) by
\begin{equation}
 \tilde{G}^{-1}(q^2) = 2\overline{m}^2(k_\Phi) +
 2Z_{\Phi ,k_\Phi}(q)q^2\; .
\end{equation}
This fixes the term in $U_{k_\Phi}$ which is quadratic in $\Phi$ to be
positive, $U_{k_\Phi}=\ol{m}^2\tr\Phi^\dagger\Phi+\ldots$. The higher
order terms in $U_{k_\Phi}$ cannot be determined in the approximation
(\ref{QCDFourFermi}) since they correspond to terms involving six or
more quark fields.  The initial value of the Yukawa coupling
corresponds to the ``quark wave function in the meson'' in eq.
(\ref{BSFact}), i.e.
\begin{equation}
 \overline{h} _{k_\Phi}(-q,q-p) = g(-q,q-p)
\end{equation}
which can be normalized with $\overline{h}_{k_\Phi}(0,0)=g(0,0)=1$.
We observe that the explicit chiral symmetry breaking from
non--vanishing current quark masses appears now in the form of a meson
source term with
\begin{equation}
  \label{AAA22}
  \jmath=2\ol{m}^2 Z_{\psi,k_\Phi}(0)
  g^{-1}(0,0)\left(
  m_{ab}+i\tilde{m}_{ab}\right)=
  2Z_{\psi,k_\Phi}\ol{m}^2
  {\rm diag}(m_u,m_d,m_s)\; .
\end{equation}
This induces a non--vanishing $\VEV{\Phi}$ and an effective quark mass
$M_q$ through the Yukawa coupling. Spontaneous chiral symmetry
breaking can be described in this language by a non--vanishing
$\VEV{\Phi}$ in the limit $\jmath\ra0$. We note that the current quark
mass $m_q$ and the constituent quark mass $M_q\sim\ol{h}\VEV{\Phi}$
are identical only for a quadratic potential
$\sim\tr\Phi^\dagger\Phi$. Because of spontaneous chiral symmetry
breaking the constituent quark mass $M_q$ can differ from zero even
for $m_q=0$. One may check that by solving the field equation for
$\Phi$ as a functional of $\ol{\psi}$, $\psi$ (with
$U_k=\ol{m}^2\tr\Phi^\dagger\Phi$) one recovers from
(\ref{EffActAnsatz}) the effective quark action
(\ref{QCDFourFermi}). Note that $\Gamma_k$ in (\ref{EffActAnsatz}) is
chirally invariant. The flow equation for $\Gamma_k$ therefore
respects chiral symmetry which leads to a considerable simplification.

At the scale $k_\Phi$ the propagator $\tilde{G}$ and the wave function
$g(-q,q-p)$ should be optimized for a most complete elimination of
terms quartic in the quark fields. In the present context we will,
however, neglect the momentum dependence of $Z_{\psi ,k}$,
$Z_{\Phi,k}$ and $\overline{h} _k$.  The mass term $\ol{m}^2$ was
found in~\cite{EW94-1} for the simple truncation (\ref{QCDFourFermi})
with $Z_\psi=1$, $m=\tilde{m}=0$ to be $\ol{m}(k_\Phi)\simeq120\MeV$.
In view of the possible large truncation errors we will take this only
as an order of magnitude estimate. Below we will consider the range
$\ol{m}(k_\Phi)=(45-120)\MeV$ for which chiral symmetry breaking can
be obtained in a two flavor model.  Furthermore, we will assume, as
usually done in large--$N_c$ computations within the NJL--model, that
$Z_{\Phi}(k_\Phi)\equiv Z_{\Phi,k_\Phi}(q=0)\ll1$.  Moreover, the
quark wave function renormalization $Z_\psi(k)\equiv Z_{\psi,k}(q=0)$
is set to one at the scale $k_\Phi$ for convenience.  For $k<k_\Phi$
we will therefore study an effective action for quarks and mesons in
the truncation
\begin{eqnarray}
 \ds{\Gm_k} &=& \ds{
 \int d^4x\Bigg\{
 Z_\psi\ol{\psi}_a i\slash{\prl}\psi^a+
 Z_\Phi\tr\left[\prl_\mu\Phi^\dagger\prl^\mu\Phi\right]+
 U_k(\Phi,\Phi^\dagger)
 }\nnn
 &+& \ds{
 \olh\,\ol{\psi}^a\left(\frac{1+\gm_5}{2}\Phi_{ab}-
 \frac{1-\gm_5}{2}(\Phi^\dagger)_{ab}\right)\psi^b
 \Bigg\} }
 \label{GammaEffective}
\end{eqnarray}
with compositeness conditions
\begin{eqnarray}
 \ds{U_{k_\Phi}(\Phi,\Phi^\dagger)} &=& \ds{
 \ol{m}^2(k_\Phi)\tr\Phi^\dagger\Phi-
 \frac{1}{2}\ol{\nu}(k_\Phi)\left(
 \det\Phi+\det\Phi^\dagger\right)}\nnn
 &+& \ds{
 \frac{1}{2}\ol{\lambda}_1\left(\tr\Phi^\dagger\Phi\right)^2+
 \frac{N-1}{4}\ol{\lambda}_2\tr
 \left(\Phi^\dagger\Phi-\frac{1}{N}\tr\Phi^\dagger\Phi\right)^2+
 \ldots}\nnn
 \ds{\olm^2(k_\Phi)} &\equiv& \ds{\frac{1}{2\tilde{G}(0)}\simeq
 (45\MeV)^2-(120\MeV)^2}\nnn
 \ds{\olh(k_\Phi)} &=&
 Z_\psi(k_\Phi)=1\nnn
 \ds{Z_\Phi(k_\Phi)} &\ll& 1\; .
 \label{CompositenessConditions}
\end{eqnarray}
As a consequence, the initial value of the renormalized Yukawa
coupling which is given by
$h(k_\Phi)=\ol{h}(k_\Phi)Z_\psi^{-1}(k_\Phi)Z_\Phi^{-1/2}(k_\Phi)$ is
large!  Note that we have included in the potential an explicit
$U_A(1)$ breaking term~\footnote{The anomaly term in the fermionic
  effective average action has been computed in~\cite{Paw96-1}.}
$\sim\olnu$ which mimics the effect of the chiral anomaly of QCD to
leading order in an expansion of the effective potential in powers of
$\Phi$. Because of the infrared stability discussed in section
\ref{InfraredStability} the precise form of the potential, i.e.~ the
values of the quartic couplings $\ol{\lambda}_i(k_\Phi)$ and so on,
will turn out to be unimportant.

We have refrained here for simplicity from considering four quark
operators with vector and pseudo--vector spin structure.  Their
inclusion is straightforward and would lead to vector and
pseudo--vector mesons in the effective action (\ref{GammaEffective}).
We will concentrate first on two flavors and consider only the two
limiting cases $\ol{\nu}=0$ and $\ol{\nu}\ra\infty$.  We also omit
first the explicit quark masses and study the chiral limit $\jmath=0$.
(The more realistic three flavor situation with a finite but
non--vanishing $\ol{\nu}$ and non--vanishing quark masses is depicted
in section \ref{PhenomenologyOfTheLinearMesonModel}.)  Because of the
positive mass term $\ol{m}^2(k_\Phi)$ one has at the scale $k_\Phi$ a
vanishing expectation value $\VEV{\Phi}=0$ (for $\jmath=0$). There is
no spontaneous chiral symmetry breaking at the compositeness scale.
This means that the mesonic bound states at $k_\Phi$ and somewhat
below are not directly connected to chiral symmetry breaking.

The question remains how chiral symmetry is broken. We will try to
answer this question by following the evolution of the effective
potential $U_k$ from $k_\Phi$ to lower scales using the exact
renormalization group method outlined in section \ref{AnExactRGE} with
the compositeness conditions (\ref{CompositenessConditions}) defining
the initial values.  In this context it is important that the
formalism for composite fields~\cite{EW94-1} also induces an infrared
cutoff in the meson propagator. The flow equations are therefore
exactly of the form (\ref{ERGE}) (except for the supertrace), with
quarks and mesons treated on an equal footing.  In fact, one would
expect that the large renormalized Yukawa coupling will rapidly drive
the scalar mass term to negative values as the IR cutoff $k$ is
lowered~\cite{Wet90-1}.  This will then finally lead to a potential
minimum away from the origin at some scale $k_{\rm \chi SB}<k_\Phi$
such that $\VEV{\Phi}\neq0$. The ultimate goal of such a procedure,
besides from establishing the onset of chiral symmetry breaking, would
be to extract phenomenological quantities, like $f_\pi$ or meson
masses, which can be computed in a straightforward manner from
$\Gamma_k$ in the IR limit $k\ra0$.

At first sight, a reliable computation of $\Gamma_{k\ra0}$ seems a
very difficult task. Without a truncation $\Gamma_k$ is described by
an infinite number of parameters (couplings, wave function
renormalizations, etc.) as can be seen if $\Gamma_k$ is expanded in
powers of fields and derivatives. For instance, the pseudoscalar and
scalar meson masses are obtained as the poles of the exact propagator,
$\lim_{k\ra0}\Gamma_k^{(2)}(q)|_{\Phi=\VEV{\Phi}}$, which receives
formally contributions from terms in $\Gamma_k$ with arbitrarily high
powers of derivatives and the expectation value $\sigma_0$.  Realistic
nonperturbative truncations of $\Gamma_k$ which reduce the problem to
a manageable size are crucial.  The remainder of this work will be
devoted to demonstrate that there may be a twofold solution to this
problem:
\begin{itemize}
\item Due to an IR fixed point structure of the flow equations in the
  symmetric regime, i.e. for $k_{\chi SB}<k<k_\Phi$, the values of
  many parameters of $\Gamma_k$ for $k\ra0$ will be approximately
  independent of their initial values at the compositeness scale
  $k_{\Phi}$. For large enough $h(k_\Phi)$ only a few relevant
  parameters ($\ol{m}^2(k_\Phi)$, $\ol{\nu}(k_\Phi)$) need to be
  computed accurately from QCD. They can alternatively be determined
  from phenomenology.
\item Physical observables like meson masses, decay constants, etc.,
  can be expanded in powers of the quark masses. This is similar to
  the way it is usually done in chiral perturbation theory
  \cite{Leu95-1}. To a given finite order of this expansion only a
  finite number of terms of a simultaneous expansion of $\Gamma_k$ in
  powers of derivatives and $\Phi$ are required if the expansion point
  is chosen properly.
\end{itemize}
In combination, these two results open the possibility for a perhaps
unexpected degree of predictive power within the linear meson model.

We wish to stress, though, that a perturbative treatment of the model
at hand, e.g., using perturbative RG techniques, cannot be expected to
yield reliable results. The renormalized Yukawa coupling is expected
to be large at the scale $k_\Phi$ and even the IR value of $h$ is
still relatively big
\begin{equation} 
  \label{IRh}
  h(k=0)=\frac{2M_q}{f_\pi}\simeq7.5 
\end{equation} 
and $h$ increases with increasing $k$. The dynamics of the linear
meson model is therefore clearly nonperturbative for all scales $k\leq
k_\Phi$.

\section{Flow equations for the linear meson model}

We will next turn to the ERGE analysis of the linear meson model which
was introduced in the last section.  To be specific, we will try to
attack the problem at hand by truncating $\Gm_k$ in such a way that it
contains all perturbatively relevant and marginal operators, i.e.
those with canonical dimensions $d_c\leq4$ in four space--time
dimensions and ignore the evolution and effects coming from operators
with $d_c>4$. The effective potential $U_k$ is a function of only four
$\csN$ invariants~\footnote{The invariant $\tau_3$ is only independent
  for $N\ge3$. For $N=2$ it can be eliminated by a suitable combination
  of $\tau_2$ and $\rho$. The additional $U_A(1)$ breaking invariant
  $\omega=i(\det\Phi-\det\Phi^\dagger)$ is $\Cc\Pc$ violating and may
  therefore appear only quadratically in $U_k$. It is straightforward
  to see that $\omega^2$ is expressible in terms of the invariants
  (\ref{Invariants}).} for $N=3$:
\begin{eqnarray}
  \label{Invariants}
  \ds{\rho} &=&
  \ds{\tr\Phi^\dagger\Phi}\nnn \ds{\tau_2} &=& \ds{
  \frac{N}{N-1}\tr\left(\Phi^\dagger\Phi- \frac{1}{N}\rho\right)^2}\nnn
  \ds{\tau_3} &=& \ds{ \tr\left(\Phi^\dagger\Phi-
  \frac{1}{N}\rho\right)^3}\nnn \ds{\xi} &=&
  \ds{\det\Phi+\det\Phi^\dagger}\; .  
\end{eqnarray}
We will expand $U_k$ as a function of these invariants around
its minimum in the absence of explicit $SU(3)$ breaking due to current
quark masses, i.e. $\rho=\rho_0\equiv N\ol{\sigma}_0^2$,
$\xi=\xi_0=2\ol{\sigma}_0^N$ and $\tau_2=\tau_3=0$ where
\begin{equation}
  \label{Sigma_0}
  \ol{\sigma}_0\equiv\frac{1}{N}\tr\VEV{\Phi}\; .
\end{equation}
Restricting $U_k$ to operators of canonical dimension $d_c\leq4$
therefore this yields in the chirally symmetric regime, i.e., for
$k_{\rm\chi SB}\leq k\leq k_\Phi$ where $\ol{\sigma}_0=0$
\begin{equation}
  \label{SymmetricPotential}
  U_k=\ol{m}^2\rho+\hal\ol{\lambda}_1\rho^2+
  \frac{N-1}{4}\ol{\lambda}_2\tau_2-
  \frac{1}{2}\ol{\nu}\xi
\end{equation}
whereas in the SSB regime for $k\leq k_{\rm\chi SB}$ we have
\begin{equation}
 U_k=\hal\olla_1(k)\left[\rho-N\ol{\sigma}_0^2(k)\right]^2+
 \frac{N-1}{4}\olla_2(k)\tau_2+
 \hal\olnu(k)\left[\ol{\sigma}_0^{N-2}(k)\rho-\xi\right]\, .
 \label{EffPotentialSSB}
\end{equation}

Before continuing to the actual beta functions for the couplings or
parameters still kept in $\Gamma_k$ it is worthwhile to pause here and
emphasize that naively (perturbatively) irrelevant operators can by no
means always be neglected. The most prominent example for this is QCD
itself. It is the very assumption of our treatment of chiral symmetry
breaking (substantiated by the results of~\cite{EW94-1}) that the
momentum dependence of the coupling constants of some six--dimensional
quark operators $(\olq q)^2$ develop poles in the $s$--channel
indicating the formation of mesonic bound states. On the other hand,
it is quite natural to assume that $\Phi^6$ or $\Phi^8$ operators are
not really necessary to understand the properties of the potential in
a neighborhood around its minimum.  Yet, truncating higher
dimensional operators does not imply the assumption that the
corresponding coupling constants are small. In fact, this could only
be expected as long as the relevant and marginal couplings are small
as well. What is required, though, is that their {\em influence} on
the evolution of those couplings kept in the truncation, for instance,
the set of equations (\ref{FlowEquationsSSB}) below, is small.  In
this context it is perhaps also interesting to note that the
truncation (\ref{GammaEffective}) includes the known one--loop beta
functions of a small coupling expansion as well as the leading order
result of the large--$N_c$ expansion of the $U_L(N)\times U_R(N)$
model~\cite{BHJ94-1}. This should provide at least some minimal
control over this truncation, even though we hope that our results are
significantly more accurate.

Inserting the truncation (\ref{GammaEffective}),
(\ref{SymmetricPotential}), (\ref{EffPotentialSSB}) into (\ref{ERGE})
reduces this functional differential equation for infinitely many
variables to a finite set of ordinary differential equations. This
yields, in particular, the beta functions for the couplings $\olla_1$,
$\olla_2$, $\olnu$ and $\olm^2$ or $\ol{\sigma}_0$. Details of the
calculation can be found in~\cite{JW95-1}. We will refrain here from
presenting the full set of flow equations but rather illustrate the
main results with a few examples. Defining dimensionless renormalized
mass, VEV and coupling constants
\begin{eqnarray}
  \ds{\eps(k)} &=& \ds{k^{-2}m^2(k)=\olm^2(k)Z_\Phi^{-1}(k)k^{-2}}\nnn
  \ds{\kappa(k)} &=& \ds{k^{2-d}N\sigma_R^2=
  Z_\Phi k^{2-d}N\ol{\sigma}_0^2}\nnn
  \ds{h^2(k)} &=& \ds{\olh^2(k)Z_\Phi^{-1}(k)Z_\psi^{-2}(k)}\nnn
  \ds{\la_i(k)} &=& \ds{\olla_i Z_\Phi^{-2}(k)\; ;\;\;\; i=1,2}\nnn
  \ds{\nu(k)} &=& \ds{\olnu(k)Z_\Phi^{-\frac{N}{2}}(k)k^{N-4}}\; .
  \label{DimensionlessCouplings}
\end{eqnarray}
one finds, e.g., for the spontaneous symmetry breaking (SSB) regime
and $\olnu=0$ 
\begin{eqnarray}
  \ds{ \frac{\prl\kappa}{\prl t} } &=& \ds{
  -(2+\eta_\Phi )\kappa + \frac{1}{16\pi^2} \Bigg\{ N^2l_1^4(0)
  +3l_1^4 (2\la_1 \kappa) }\nnn &+& \ds{ (N^2-1)\left[
  1+\frac{\la_2}{\la_1}\right] l_1^4 (\la_2\kappa)-4N_c
  \frac{h^2}{\la_1} l_{1}^{(F)4} (\frac{1}{N}h^2 \kappa) \Bigg\} }\nnn
  \ds{\frac{\prl\la_1}{\prl t} } &=& \ds{ 2\eta_\Phi \la_1
  +\frac{1}{16\pi^2} \Bigg\{ N^2\la_1^2 l_2^4(0) +9\la_1^2 l_2^4
  (2\la_1\kappa) }\nnn &+& \ds{ (N^2-1)\left[\la_1 +\la_2\right]^2
  l_2^4 (\la_2\kappa)-4\frac{N_c}{N}h^4 l_{2}^{(F)4}
  (\frac{1}{N}h^2\kappa) \Bigg\} \label{FlowEquationsSSB} }\\[2mm]
  \ds{\frac{\prl\la_2}{\prl t} } &=& \ds{ 2\eta_\Phi \la_2
  +\frac{1}{16\pi^2} \Bigg\{ \frac{N^2}{4}\la_2^2 l_2^4(0) +
  \frac{9}{4}(N^2-4)\la_2^2 l_2^4 (\la_2\kappa) }\nnn &-& \ds{ \hal
  N^2 \la_2^2 l_{1,1}^4(0,\la_2\kappa) + 3[\la_2+4\la_1]\la_2
  l_{1,1}^4(2\la_1\kappa,\la_2\kappa) }\nnn &-& \ds{ 8\frac{N_c}{N}h^4
  l_{2}^{(F)4} (\frac{1}{N}h^2\kappa)\Bigg\} }\nnn \displaystyle{
  \frac{\partial}{\partial t} h^2} &=& \displaystyle{ \left[
  d-4+2\eta_\psi +\eta_\varphi\right] h^2 -\frac{4}{N}v_d h^4 \left\{
  N^2l_{1,1}^{(FB)d} (\frac{1}{N}\kappa h^2,\epsilon;
  \eta_\psi,\eta_\varphi) \right. }\nonumber\vspace{.2cm}\\ &-&
  \displaystyle{\left.  (N^2-1)l_{1,1}^{(FB)d} (\frac{1}{N}\kappa
  h^2,\epsilon+\kappa\lambda_2; \eta_\psi,\eta_\varphi)
  -l_{1,1}^{(FB)d} (\frac{1}{N}\kappa h^2,\epsilon+2\kappa\lambda_1;
  \eta_\psi,\eta_\varphi) \right\}\nonumber } 
\end{eqnarray} 
Here $\eta_\Phi=-\prl_t\ln Z_\Phi$, $\eta_\psi=-\prl_t\ln Z_\psi$ are
the meson and quark anomalous dimensions, respectively~\cite{JW95-1}.
The symbols $l_n^4$, $l_{n_1,n_2}^4$ and $l_n^{(F)4}$ denote mass
threshold functions. A typical example is
\begin{equation}
 \label{AAA85}
 l_n^4(w)=8n\pi^2
 k^{2n-4}\int\frac{d^4q}{(2\pi)^4}
 \frac{\prl_t(Z_\Phi^{-1}R_k(q^2))}
 {\left[ P(q^2)+k^2w\right]^{n+1}}
\end{equation}
with $P(q^2)=q^2+Z_\Phi^{-1}R_k(q^2)$. These functions decrease
monotonically with their arguments $w$ and decay $\sim w^{-(n+1)}$ for
$w\gg1$. Since the arguments $w$ are generally the (dimensionless)
squared masses of the model, the main effect of the threshold
functions is to cut off quantum fluctuations of particles with masses
$M^2\gg k^2$. Once the scale $k$ is changed below a certain mass
threshold, the corresponding particle no longer contributes to the
evolution of the couplings and decouples smoothly.  

Within our truncation the beta functions (\ref{FlowEquationsSSB}) for
the dimensionless couplings look almost the same as in one--loop
perturbation theory. There are, however, two major new ingredients
which are crucial for our approach: First, there is a new equation for
the running of the mass term in the symmetric regime or for the
running of the potential minimum in the regime with spontaneous
symmetry breaking. This equation is related to the quadratic
divergence of the mass term in perturbation theory and does not appear
in the Callan--Symanzik~\cite{CS70-1} or
Coleman--Weinberg~\cite{CW73-1} treatment of the renormalization
group. Obviously, this equation is the key for a study of the onset of
spontaneous chiral symmetry breaking as $k$ is lowered from $k_\Phi$
to zero.  Second, the most important nonperturbative ingredient in the
flow equations for the dimensionless Yukawa and scalar couplings is
the appearance of effective mass threshold functions like
(\ref{AAA85}) which account for the decoupling of modes with masses
larger than $k$.  Their form is different for the symmetric regime
(massless fermions, massive scalars) or the regime with spontaneous
symmetry breaking (massive fermions, massless Goldstone bosons).
Without the inclusion of the threshold effects the running of the
couplings would never stop and no sensible limit $k\rightarrow0$ could
be obtained due to unphysical infrared divergences. The threshold
functions are not arbitrary but have to be computed carefully. The
mass terms appearing in these functions involve the dimensionless
couplings. Expanding the threshold functions in powers of the mass
terms (or the dimensionless couplings) makes their non--perturbative
content immediately visible.  (It is the presence of threshold
functions which explains why one--loop type formulae could be used for
a necessarily nonperturbative computation of critical exponents in
three--dimensional scalar theories \cite{TW94-1}.)

\section{The chiral anomaly and the $O(4)$--model}
\label{ChiralAnomaly}

We have seen how the mass threshold functions in the flow equations
describe the decoupling of heavy modes from the evolution of
$\Gamma_k$ as the IR cutoff $k$ is lowered. In the chiral limit with
two massless quark flavors ($N=2$) the pions are the massless
Goldstone bosons whereas all other mesons have masses larger than
$k_\Phi$ and are therefore practically decoupled already at the scale
$k_\Phi$. The effect of the physical pion mass of
$m_\pi\simeq140\MeV$, or equivalently of the two small but
non--vanishing current quark masses, can then easily be mimicked by
stopping the flow of $\Gamma_k$ at $k=m_\pi$ by hand.  This situation
changes significantly once the strange quark is included. Now the
$\eta$ and the four $K$ mesons appear as additional massless Goldstone
modes in the spectrum. They would artificially drive the running of
$\Gamma_k$ at scales $m_\pi\lta k\lta500\MeV$ where they should
already be decoupled because of their physical masses. It is therefore
advisable to focus on the two flavor case $N=2$ as long as the chiral
limit of vanishing current quark masses is considered.

It is straightforward to obtain an estimate of the (renormalized)
coupling $\nu$ parameterizing the explicit $U_A(1)$ breaking due to
the chiral anomaly. From (\ref{EffPotentialSSB}) we find
\begin{equation}
  \label{AAA01}
  m_{\eta^\prime}^2=\frac{N}{2}\nu\ol{\sigma}_0^{N-2}\simeq1\GeV
\end{equation}
which translates for $N=2$ into
\begin{equation}
  \label{AAA02}
  \nu(k\ra0)\simeq1\GeV\; .
\end{equation}
This suggests that $\nu\ra\infty$ can be considered as a realistic
limit.  An important simplification occurs for $N=2$ and
$\nu\ra\infty$, related to the fact that for $N=2$ the chiral group
$SU_L(2)\times SU_R(2)$ is (locally) isomorphic to $O(4)$. Thus, the
complex $(\bf 2,\bf 2)$ representation $\Phi$ of $SU_L(2)\times
SU_R(2)$ may be decomposed into two vector representations,
$(\si,\pi^k)$ and $(\eta^\prime,a^k)$ of $O(4)$:
\begin{equation}
 \Phi=\hal\left(\si-i\eta^\prime\right)+
 \hal\left( a^k+i\pi^k\right)\tau_k \; .
\end{equation}
For $\nu\ra\infty$ the masses of the $\eta^\prime$ and the $a^k$
diverge and these particles decouple. We are then left with the
original $O(4)$ symmetric linear $\si$--model of Gell--Mann and Levy
\cite{GML60-1} coupled to quarks. The flow equations of this model
have been derived previously~\cite{Wet90-1,Wet91-1} for the truncation
of the effective action used here. For mere comparison we also
consider the opposite limit $\nu\ra0$. Here the $\eta^\prime$ meson
becomes an additional Goldstone boson in the chiral limit which
suffers from the same problem as the $K$ and the $\eta$ in the case
$N=3$.  Hence, we may compare the results for two different
approximate limits of the effects of the chiral anomaly:
\begin{itemize}
\item the $O(4)$ model corresponding to $N=2$ and $\nu\ra\infty$
\item the $U_L(2)\times U_R(2)$ model corresponding to $N=2$ and
  $\nu=0$.
\end{itemize}
For the reasons given above we expect the first situation to be closer
to reality. In this case we may imagine that the fluctuations of
kaons, $\eta$ $\eta^\prime$ and the scalar mesons (as well as vector
and pseudovector mesons) have been integrated out in order to obtain
the initial values of $\Gamma_{k_\Phi}$ --- in close analogy to the
integration of the gluons for the effective quark action
$\Gamma_{k_p}[\psi]$ discussed in section
\ref{ASemiQuantitativePicture}. We will keep the initial values of the
couplings $\ol{m}^2(k_\Phi)$, $\ol{\lambda}_1(k_\Phi)$ and
$Z_\Phi(k_\Phi)$ as free parameters. Our results should be
quantitatively accurate to the extent to which the local polynomial
truncation is a good approximation.

\section{Infrared stability}
\label{InfraredStability}

\begin{figure}
\unitlength1.0cm
\begin{picture}(13.,8.)
\put(1.4,5.){\bf $\ds{\frac{m,\sigma_R}{\MeV}}$}
\put(8.5,0.5){\bf $k/\MeV$}
\put(6.5,3.5){\bf $\sigma_R$}
\put(12.0,5.){\bf $m$}
\put(0.2,-11.5){
\epsfysize=22.cm
\epsffile{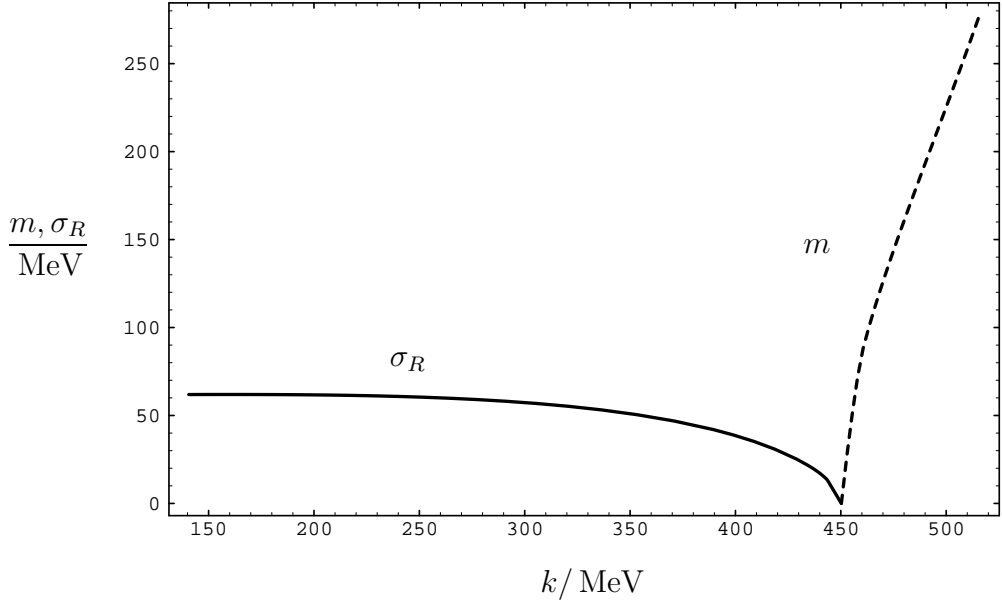}
}
\end{picture}
\caption{Evolution of the renormalized mass $m$ in the symmetric
  regime (dashed line) and the vacuum expectation value
  $\sigma_R=Z_\Phi^{1/2}\ol{\sigma}_0$ of the scalar field in the SSB
  regime (solid line) as functions of $k$ for the $U_L(2)\times
  U_R(2)$ model. Initial values are $\la_i(k_\Phi)=0$,
  $h^2(k_\Phi)=300$ and $\ol{m}(k_\Phi)=63\MeV$.}
\label{Fig1} 
\end{figure}

Eqs.~(\ref{FlowEquationsSSB}) and the corresponding set of flow
equations for the symmetric regime constitute a coupled system of
ordinary differential equations which can be integrated numerically.
Similar equations obtain for the $O(4)$ model where $N=2$ and the
coupling $\lambda_2$ is absent.  The most important result is that
chiral symmetry breaking indeed occurs for a wide range of initial
values of the parameters including the presumably realistic case of
large renormalized Yukawa coupling and a bare mass $\olm(k_\Phi)$ of
order $100\MeV$. A typical evolution of the renormalized mass $m(k)$
is plotted in figure \ref{Fig1}.
Driven by the strong Yukawa coupling, $m$ decreases rapidly and goes
through zero at a scale $k_{\chi{\rm SB}}$ not far below $k_\Phi$.
Here the system enters the SSB regime and a non--vanishing
(renormalized) VEV $\sigma_R$ for the meson field $\Phi$ develops. The
evolution of $\sigma_R(k)$ turns out to be reasonably stable already
before scales $k\simeq m_\pi$ where the evolution is stopped. We take
this result as an indication that our truncation of the effective
action $\Gm_k$ leads at least qualitatively to a satisfactory
description of chiral symmetry breaking. The reason for the relative
stability of the IR behavior of the VEV (and all other couplings) is
that the quarks acquire a constituent mass
$M_q=h\sigma_R\simeq350\MeV$ in the SSB regime. As a consequence they
decouple once $k$ becomes smaller than $M_q$ and the evolution is then
dominantly driven by the massless Goldstone bosons.  This is also
important in view of potential confinement effects expected to become
important for the quarks for $k$ around $\La_{\rm QCD}\simeq200\MeV$.
Since confinement is not explicitly included in our truncation of
$\Gamma_k$, one might be worried that such effects could spoil our
results completely. Yet, as discussed in some more detail in section
\ref{ASemiQuantitativePicture}, only the colored quarks should feel
confinement and they are no longer important for the evolution of
the meson couplings for $k$ around $200\MeV$.  One might therefore hope
that a precise treatment of confinement is not crucial for this
approach to chiral symmetry breaking.

\begin{figure}
\unitlength1.0cm
\begin{picture}(13.,9.)
\put(1.5,5.){\bf $\ds{\frac{f_\pi}{\MeV}}$}
\put(8.5,0.5){\bf $\tilde{\epsilon}_0$}
\put(11.0,5.5){\bf $O(4)$}
\put(6.0,2.5){\bf $U_L(2)\times U_R(2)$}
\put(0.2,-11.5){
\epsfysize=22.cm
\epsffile{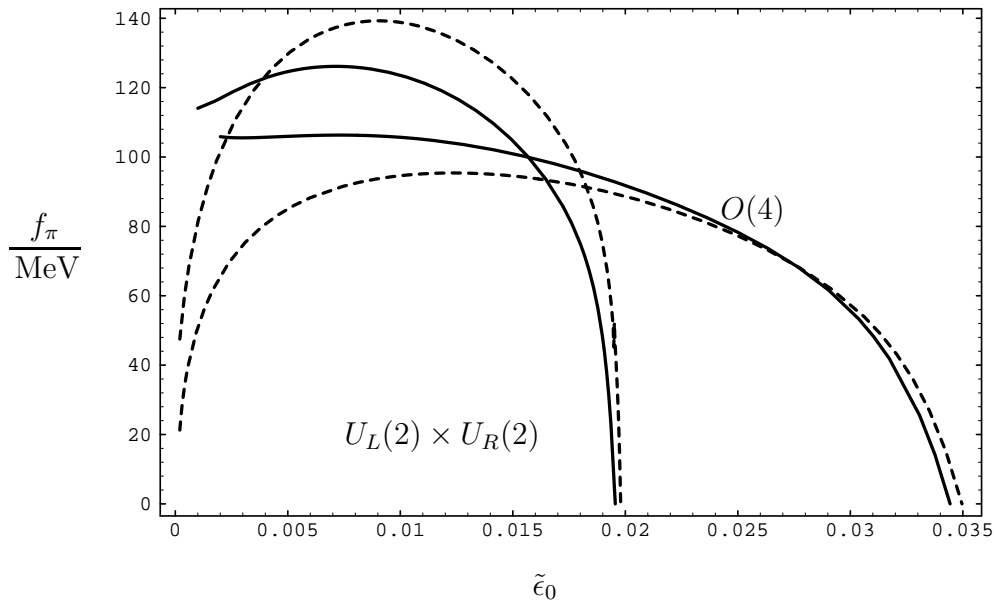}
}
\end{picture}
\caption{The pion decay constant $f_\pi$ as a function of $\teps_0$
for $k_\Phi=630\MeV$, $\la_1(k_\Phi)=\la_2(k_\Phi)=0$ and
$h^2(k_\Phi)=300$ (solid line) as well as $h^2(k_\Phi)=10^4$ (dashed
line).}
\label{Fig8}
\end{figure}

Most importantly, one finds that the system of flow equations exhibits
an IR fixed point in the symmetric phase. As already pointed out one
expects $Z_\Phi$ to be rather small at the compositeness scale $k_\Phi$.
In turn, one may assume that, at least for the initial range of
running in the symmetric regime the mass parameter $\eps\sim
Z_\Phi^{-1}$ is large. This means, in particular, that all threshold
functions with arguments $\sim\eps$ may be neglected in this regime.
As a consequence, the flow equations simplify considerably. We find,
for instance, for the $U_L(2)\times U_R(2)$ model 
\begin{eqnarray}
  \label{AAA08}
  \ds{\prl_t\teps} &\equiv& \ds{\prl_t\frac{\eps}{h^2}\simeq
    -2\teps+\frac{N_c}{4\pi^2}}\nnn \ds{\prl_t\tla_1} &\equiv&
  \ds{\prl_t\frac{\la_1}{h^2}\simeq
    \frac{N_c}{4\pi^2}h^2\left[\hal\tla_1-\frac{1}{N}\right]}\nnn
  \ds{\prl_t\tla_2} &\equiv& \ds{\prl_t\frac{\la_2}{h^2}\simeq
    \frac{N_c}{4\pi^2}h^2\left[\hal\tla_2-\frac{2}{N}\right]}\\[2mm]
  \ds{\prl_t h^2} &\simeq& \ds{\frac{N_c}{8\pi^2}h^4}\nnn
  \ds{\eta_\Phi} &\simeq& \ds{ \frac{N_c}{8\pi^2}h^2}\nnn
  \ds{\eta_\psi} &\simeq& 0\nonumber\; .
\end{eqnarray} 
This system possesses an attractive IR fixed point for the
quartic scalar self interactions 
\begin{equation}
  \label{AAA10}
  \tla_{1*}=\hal\tla_{2*}=\frac{2}{N}\; .  
\end{equation}
Furthermore it is exactly soluble~\cite{JW95-1}.  Because of the
strong Yukawa coupling the quartic couplings $\tla_1$ and $\tla_2$
generally approach their fixed point values rapidly, long before the
systems enters the broken phase ($\eps\ra0$) and the approximation of
large $\eps$ breaks down.  In addition, for large $h^2(k_\Phi)$ the
value of the Yukawa coupling at the scale $k_s$ where $\epsilon$
vanishes (or becomes small) only depends on the initial value
\begin{equation}
  \teps_0\equiv\frac{\eps(k_\Phi)}{h^2(k_\Phi)}=
  \frac{\olm^2(k_\Phi)}{k_\Phi^2}\; .
\end{equation}
Hence, the system is approximately independent in the IR upon the
initial values of $\la_1$, $\la_2$ and $h^2$, the only ``relevant''
parameter being~\footnote{Once quark masses and a proper treatment of
  the chiral anomaly are included for $N=3$ one expects that $m_q$ and
  $\nu$ are additional relevant parameter. Their values may be fixed
  by using the masses of $\pi$, $K$ and $\eta^\prime$ as
  phenomenological input.} $\teps_0$. In other words, the effective
action looses almost all its ``memory'' in the far IR of where in the
UV it came from. This feature of the flow equations leads to a perhaps
surprising degree of predictive power.  In addition, also the
dependence of $f_\pi=2\sigma_R(k\ra0)$ on $\teps_0$ is not very strong
for a large range in $\tilde{\epsilon}_0$, as shown in figure
\ref{Fig8}.
The relevant parameter $\teps_0$ can be fixed by using the constituent
quark mass $M_q\equiv(h\sigma_R)(k=0)\simeq350\MeV$ as a
phenomenological input.  One obtains for the $O(4)$ model
\begin{equation}
 \ds{\teps_0} \simeq \ds{0.02}\; .
\end{equation}
The resulting value for the decay constant is
\begin{equation}
  \label{AAA52}
  f_\pi=(91-100)\MeV
\end{equation}
for $h^2(k_\Phi)=10^4-300$. It is striking that this comes close to
the real value $f_\pi=92.4\MeV$ but we expect that the uncertainty in
the determination of the compositeness scale $k_\Phi$ and the
truncation errors exceed the influence of the variation of
$h^2(k_\Phi)$.  We have furthermore used this result for an estimate
of the chiral condensate: 
\begin{equation}
 \VEV{\ol{\psi}\psi}\equiv
 -\tilde{\epsilon}_0 f_\pi k_\Phi^2
 Z_\Phi^{-1/2}(k=0)A\simeq-(195\MeV)^3 
\end{equation} 
where the factor $A\simeq1.7$ accounts for the change of the
normalization scale of $\VEV{\ol{\psi}\psi}$ from $k_\Phi$ to the
commonly used value $1\GeV$. Our value is in reasonable agreement with
results from sum rules~\cite{JM95-1}.  This result is non--trivial
since not only $k_\Phi$ and $f_\pi$ enter but also
$\tilde{\epsilon}_0$ and the IR value $Z_\Phi(0)$.  Integrating
(\ref{AAA08}) for $\eta_\Phi$ one finds
\begin{equation}
  \label{AAA09}
  Z_\Phi(k)=Z_\Phi(k_\Phi)+
  \frac{N_c}{8\pi^2}\ln\frac{k_\Phi}{k}\; .
\end{equation}
Thus $Z_\Phi(k)$ will indeed be practically independent of its initial
value $Z_\Phi(k_\Phi)$ already after some running as long as
$Z_\Phi(k_\Phi)$ is small compared to $0.01$.

The alert reader may have noticed that the beta functions
(\ref{AAA08}) correspond exactly to those obtained in the
one--quark--loop approximation or, in other words, to the leading
order in the large--$N_c$ expansion for the Nambu--Jona-Lasinio
model~\cite{BHJ94-1}. The fixed point (\ref{AAA10}) is then nothing
but the large--$N_c$ boundary condition on the evolution of
$\lambda_1$ and $\lambda_2$ in this model. Yet, we wish to stress that
nowhere we have made the assumption that $N_c$ is a large number.  On
the contrary, the physical value $N_c=3$ suggests that the
large--$N_c$ expansion should a priori only be trusted on a
quantitatively rather crude level. The reason why we expect
(\ref{AAA08}) to nevertheless give rather reliable results is based on
the fact that for small $Z_\Phi(k_\Phi)$ all (renormalized) meson
masses are much larger than the scale $k$ for the initial part of the
running.  This implies that the mesons are effectively decoupled and
their contribution to the beta functions is negligible leading to the
one--quark--loop approximation. Yet, already after some relatively
short period of running the renormalized meson masses approach zero
and our approximation of neglecting mesonic threshold functions breaks
down.  Hence, the one--quark--loop approximation is reasonable only for
scales close to $k_\Phi$ but is bound to fail around $k_{\chi{\rm
    SB}}$ and in the SSB regime. What is important in our context is
not the numerical value of the partial fixed points (\ref{AAA10}) but
rather their mere existence and the presence of a large coupling $h^2$
driving the $\tilde{\lambda}_i$ fast towards them. This is enough for
the IR values of all couplings to become almost independent of the
initial values $\lambda_i(k_\Phi)$. Similar features of IR stability
are expected if the truncation is enlarged, for instance, to a more
general form of the effective potential $U_{k_\Phi}$.

\section{Phenomenology of the linear meson model for three flavors}
\label{PhenomenologyOfTheLinearMesonModel}

In the preceding sections we have presented an investigation of
the linear meson model subject to some simplifications and
approximations:
\begin{itemize}
\item We have restricted ourselves to the case of two light quark
  flavors, $N=2$, whereas it is known from chiral perturbation
  theory~\cite{Leu95-1} that the strange quark can also be considered
  as light.
\item Explicit chiral symmetry breaking due to current quark masses
  has been neglected. For $N=2$ its main effect is to give the pions a
  non--vanishing mass which is easily mimicked by stopping the RG
  evolution around $k=m_\pi$ by hand. For $N=3$ this would be rather
  inaccurate, since the kaons and the $\eta$ have masses quite
  different from the pions and should therefore decouple much
  earlier. This necessitates an explicit consideration of the source
  term $\sim\jmath$.
\item We have treated the effects of the chiral anomaly only for two
  limiting cases: $\nu=0$ and $\nu\ra\infty$. In reality $\nu$ is,
  however, large but finite.
\item The IR--cut--off effective action $\Gamma_k$ was truncated in
  such a way that only operators of canonical dimension $d_c\le4$ were
  kept.
\end{itemize}
It is, of course, necessary to investigate the effects of these
simplifications systematically. It is straightforward to relax the
first three points, i.e. to include three massive quark flavors and
allow for a non--vanishing and finite $\nu$. A systematic study of the
truncation errors is more involved. Even though it is possible to
include additional terms in $\Gamma_k$ it would be desirable to have a
guiding principle at hand, which would allow to determine the
invariants which one should keep in a truncation of $\Gamma_k$ and
those which can be neglected.  Clearly, such a principle will depend
on the kind of physical questions one poses. In the following we will
apply three different criteria for this purpose:
\begin{enumerate}
\item Phenomenological observables like meson masses or decay
  constants can be expanded systematically in powers of (current)
  quark masses in the linear meson model in a similar way as it is
  usually done in chiral perturbation theory. This can be used to
  determine the operators of $\Gamma_k$ which are necessary to compute
  an observable to a given order in $m_q$.
\item We have assumed so far that all other low--energy degrees of
  freedom of strong interaction physics like vector and axial--vector
  mesons have been integrated out. Carrying out this procedure
  explicitly in some cases induces relatively large values for several
  non--renormalizable interactions in the linear meson model. They
  have to be taken into account, at least for comparison with
  phenomenology.
\item Even though the range of evolution from $k_\Phi$ to $k=m_\pi$ is
  not too large ($m_\pi/k_\Phi\simeq0.2$), it is likely that higher
  dimensional operators are dimensionally suppressed at $k=m_\pi$ as
  long as their anomalous dimension is not very large. This is related
  to the ``triviality'' of the linear quark meson system, i.e. the
  existence of the interaction free ``Gaussian'' fixed point. We will
  therefore assume that of two operators contributing to a physical
  observable at the same order in the quark mass expansion, the lower
  dimensional one dominates, provided the other one does not receive
  large corrections from other degrees of freedom assumed to be
  integrated out. This can also be understood as an expansion in
  powers of the chiral condensate $\sim\ol{\sigma}_0$.
\end{enumerate}
We will sketch in the following how these criteria can be applied to
the linear meson model. The details of this procedure can be found
in~\cite{JW96-1}.

In order to perform a quark mass expansion of meson observables it
is useful to decompose $\Phi$ with respect to the vector--like
$SU_V(3)$ symmetry
\begin{equation}
  \label{AAA24}
  \Phi=\ol{\sigma}_0+
  \frac{1}{\sqrt{2}}\left(
  i\Phi_p+\frac{i}{\sqrt{3}}\chi_p+
  \Phi_s+\frac{1}{\sqrt{3}}\chi_s\right)\; .
\end{equation}
For $N=3$ the hermitean traceless $3\times3$ matrices $\Phi_p$ and
$\Phi_s$ represent (up to mixing effects) the pseudoscalar and scalar
octets, respectively. The real fields $\chi_p$ and $\chi_s$ are
associated with the parity odd and even singlets corresponding to the
$\eta^\prime$ meson and the $\sigma$--resonance, respectively. The
constant $\ol{\sigma}_0$ is chosen as the VEV of $\Phi$ in the limit of
equal quark masses $m_q=(m_u+m_d+m_s)/3$. Differences between the
strange, down and up, quark masses imply
\begin{equation}
  \label{AAA25}
  \VEV{\Phi_s}=\frac{1}{\sqrt{2Z_h}}\left(
  w\lambda_3-\sqrt{3}v\lambda_8\right)\; .
\end{equation}
Here $\lambda_3$ and $\lambda_8$ denote the two diagonal Gell--Mann
matrices. The parameters
$w$ and $v$ correspond to the explicit isospin and $SU_V(3)$ violation
in the pseudoscalar meson decay constants
\begin{eqnarray}
  \label{AAA26}
  \left(\frac{Z_m}{Z_h}\right)^{\frac{1}{2}}w &=& \ds{
  \ol{f}_{K^\pm}-\ol{f}_{K^o}}\nnn
  \left(\frac{Z_m}{Z_h}\right)^{\frac{1}{2}}v &=& \ds{
  \frac{1}{3}\left(
  \ol{f}_{K^\pm}+\ol{f}_{K^o}-2\ol{f}_\pi\right) }\; .
\end{eqnarray}
Here $Z_m$ and $Z_h$ denote the wave functions renormalization
constants of $\Phi_p$ and $\Phi_s$, respectively.  The $\ol{f}_i$ are
related to the physical decay constants $f_i$ by additional wave
function renormalization effects which will be discussed below.

It is convenient to expand the effective potential around
$\Phi=\ol{\sigma}_0$ or $\rho=3\ol{\sigma}_0^2$,
$\xi=2\ol{\sigma}_0^3$ and $\tau_2=\tau_3=0$:
\begin{eqnarray}
  \label{AAA27}
  \ds{U(\rho,\tau_2,\tau_3,\xi) } &=& \ds{
  \ol{m}_g^2\left(\rho-\rho_0\right)
  -\hal\ol{\nu}\left[\xi-\xi_0-\ol{\si}_0(\rho-\rho_0)\right]}\nnn
  &+& \ds{
  \hal\ol{\la}_1\left(\rho-\rho_0\right)^2+\hal\ol{\la}_2\tau_2+
  \hal\ol{\la}_3\tau_3}  \\[2mm]
  &+& \ds{
  \hal\ol{\beta}_1\left(\rho-\rho_0\right)\left(\xi-\xi_0\right)+
  \hal\ol{\beta}_2\left(\rho-\rho_0\right)\tau_2}\nnn
  &+& \ds{
  \hal\ol{\beta}_3\left(\xi-\xi_0\right)\tau_2+
  \hal\ol{\beta}_4\left(\xi-\xi_0\right)^2+\ldots\; .
  \nonumber}
\end{eqnarray}
The bare meson mass squared matrix is obtained from the second
derivatives of $U$ with respect to the components of the fields
$\Phi_s$, $\Phi_p$, $\chi_s$ and $\chi_p$ evaluated at the minimum of
$U_k-\frac{1}{2}\tr\jmath(\Phi+\Phi^\dagger)$.  Decomposing the chiral
invariants $\rho$, $\tau_2$, $\tau_3$ and $\xi$ with respect to the
$SU_V(3)$ multiplets $\Phi_s$, $\Phi_p$, $\chi_s$ and $\chi_p$ and
taking into account that the only non--vanishing VEV is
$\VEV{\Phi_s}\sim\Oc(m_q)$ it is straightforward to see that the
potential (\ref{AAA27}) contains all terms which are necessary and
sufficient for a computation of the bare pseudoscalar meson masses to
second order and the scalar octet and pseudoscalar singlet masses to
linear order in the quark masses~\footnote{If one counts the effects
  of the chiral anomaly as being of $\Oc(m_q)$, eq.~(\ref{AAA27}) is
  also sufficient for the bare pseudoscalar singlet or (up to mixing)
  the $\eta^\prime$ meson mass to quadratic order in $m_q$.}.

In a similar way one can see that modifications of the minimal kinetic
term for $\Phi$ beyond the approximation (\ref{GammaEffective}) are
necessary for a determination of the physical meson masses to the
desired order in $m_q$. There are various two--derivative terms all
contributing to the same order in the quark mass expansion to, e.g.,
meson masses. As argued above we will, however, assume that the lowest
dimensional ones dominate. The ones relevant here are given by
\begin{equation}\begin{array}{rcl}
 \ds{\Lc_{\rm kin}} &=& \ds{
 Z_\Phi\Tr\prl^\mu\Phi^\dagger\prl_\mu\Phi}\nnn
 &+& \ds{\frac{1}{2}U_\Phi
 \eps^{a_1a_2a_3}\eps^{b_1b_2b_3}
 \left(\Phi_{a_1b_1}\prl^\mu\Phi_{a_2b_2}\prl_\mu\Phi_{a_3b_3}+
 \Phi^\dagger_{a_1b_1}\prl^\mu\Phi^\dagger_{a_2b_2}
 \prl_\mu\Phi^\dagger_{a_3b_3}
 \right)}\nnn
 &-& \ds{\frac{1}{8}X_\Phi^-\Big\{\Tr\left(
 \Phi^\dagger\prl_\mu\Phi-\prl_\mu\Phi^\dagger\Phi\right)
 \left(\Phi^\dagger\prl^\mu\Phi-\prl^\mu\Phi^\dagger\Phi\right)}\nnn
 && \ds{\hspace{8.5mm}+
 \Tr\left(\Phi\prl_\mu\Phi^\dagger-\prl_\mu\Phi\Phi^\dagger\right)
 \left(\Phi\prl^\mu\Phi^\dagger-\prl^\mu\Phi\Phi^\dagger\right)
 \Big\} }\nnn
 &-& \ds{\frac{1}{8}X_\Phi^+\Big\{\Tr\left(
 \Phi^\dagger\prl_\mu\Phi+\prl_\mu\Phi^\dagger\Phi\right)
 \left(\Phi^\dagger\prl^\mu\Phi+\prl^\mu\Phi^\dagger\Phi\right)}\nnn
 && \ds{\hspace{8.5mm}+
 \Tr\left(\Phi\prl_\mu\Phi^\dagger+\prl_\mu\Phi\Phi^\dagger\right)
 \left(\Phi\prl^\mu\Phi^\dagger+\prl^\mu\Phi\Phi^\dagger\right)
 \Big\} +\ldots } \; .
 \label{Lkin}
\end{array}\end{equation}
These modifications of the kinetic term induce different wave function
renormalization constants $Z_h$, $Z_m$, $Z_s$ and $Z_p$ for the four
multiplets $\Phi_s$, $\Phi_p$, $\chi_s$ and $\chi_p$, respectively,
e.g.
\begin{eqnarray}
  \label{AAA29a}
  \ds{Z_m} &=& \ds{
  Z_\Phi+U_\Phi\ol{\sigma}_0+X_\Phi^-\ol{\sigma}_0^2}\nnn
  \ds{Z_p} &=& \ds{
  Z_\Phi-2U_\Phi\ol{\sigma}_0+X_\Phi^-\ol{\sigma}_0^2}\nnn
  \ds{Z_h} &=& \ds{
  Z_\Phi-U_\Phi\ol{\sigma}_0-X_\Phi^+\ol{\sigma}_0^2}\; .
\end{eqnarray}
Furthermore, they lead to a split in the wave function
renormalizations for the individual members of each multiplet:
\begin{eqnarray}
  \label{AAA29b}
  \ds{Z_\pi} &=& \ds{1-\omega_m v}\nnn
  \ds{Z_{K^\pm}} &=& \ds{
  1+\frac{1}{2}\omega_m\left( v+w\right)}\nnn
  \ds{Z_{K^o}} &=& \ds{
  1+\frac{1}{2}\omega_m\left( v-w\right)}\nnn
  \ds{Z_8} &=& \ds{1+\omega_m v}
\end{eqnarray}
where
\begin{eqnarray}
  \label{AAA29}
  \ds{\omega_m} &=& \ds{
  \left( X_\Phi^-\ol{\sigma}_0-U_\Phi\right)
  Z_h^{-1/2}Z_m^{-1}}
\end{eqnarray}
The physical meson masses and decay constants are then
given by
\begin{equation}
  \label{AAA29c}
  M_\pi^2=\hat{M}_\pi^2 Z_\pi^{-1}Z_m^{-1}\; ;\;\;\;
  f_\pi=\ol{f}_\pi Z_\pi^{1/2}\; ;\;\;\;
  {\rm etc.}\; ,
\end{equation}
where the $\hat{M}_i^2$ are the bare meson mass terms given by the
eigenvalues of the matrix of second derivatives of the potential $U$.
For the pseudoscalar octet the relations (\ref{AAA29c}) with
(\ref{AAA29b}) contain all contributions to the meson masses to second
order and the decay constants to linear order in the quark mass
expansion.

\begin{figure}
\unitlength1.0cm
\begin{picture}(13.,8.5)
\put(.7,5.){\bf $\ds{\frac{M_\eta}{\MeV}}$}
\put(7.5,0.8){\bf $\om_m v$}

\put(7.9,6.4){$\frac{Z_p}{Z_m}=1.3$}
\put(5.7,3.0){$\frac{Z_p}{Z_m}=0.7$}

\put(-0.8,-11.5){
\epsfysize=22.cm
\epsffile{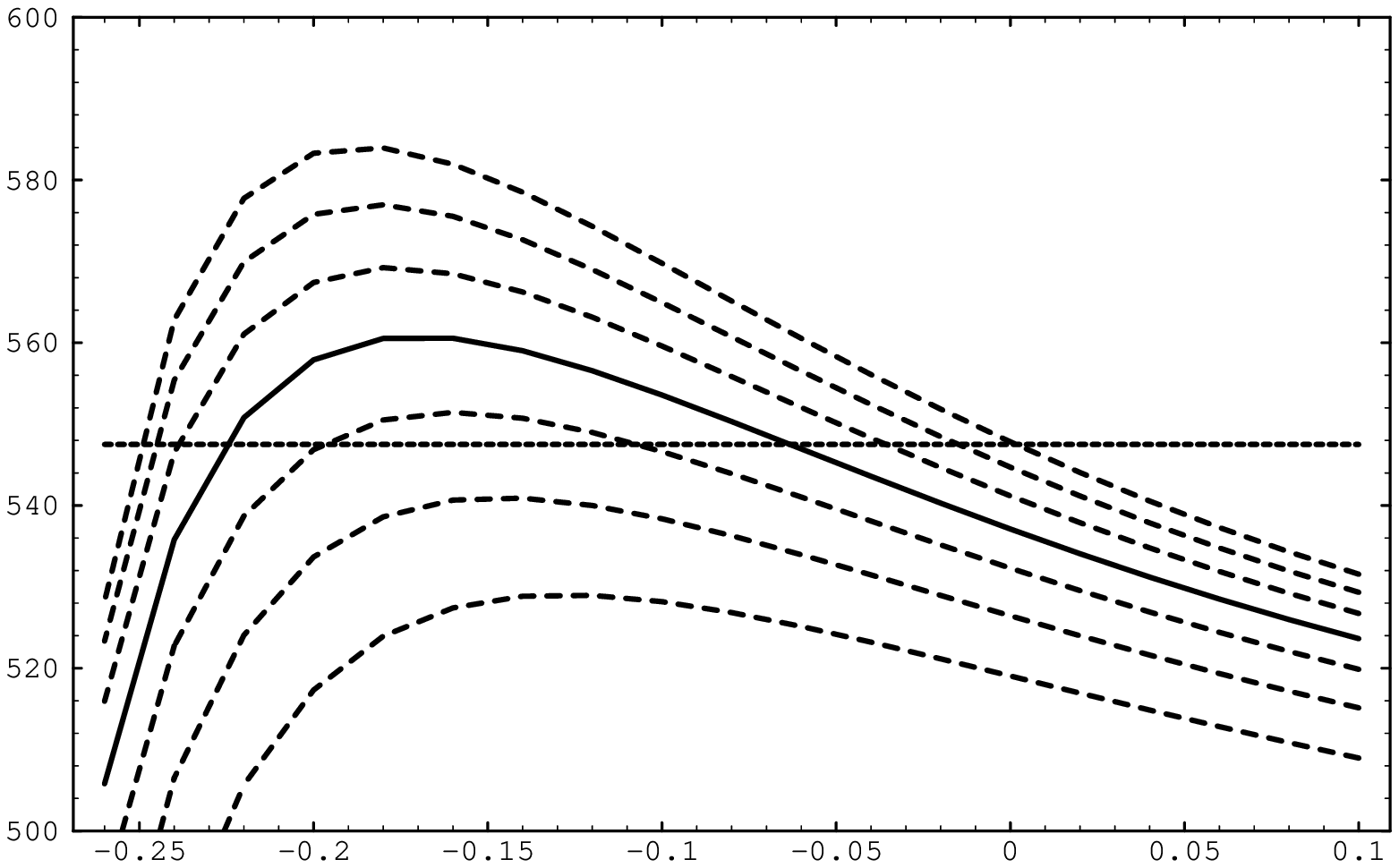}
}
\end{picture}
\caption{\footnotesize The plot shows $M_\eta$ as a function of
  $\om_m v$ for various values of $Z_p/Z_m$.  The solid line
  corresponds to $Z_p/Z_m=1$ and the difference in $Z_p/Z_m$ between
  two adjacent lines is $0.1$. The horizontal dotted line indicates
  the experimental value $M_\eta\simeq547.5\MeV$.}
\label{Plot1} 
\end{figure}

\begin{figure}
\unitlength1.0cm
\begin{picture}(13.,15.)
\put(1.2,4.6){\bf $\ds{\frac{f_{\eta^\prime}}{f_\pi}}$}
\put(1.2,11.7){\bf $\ds{\frac{f_{\eta}}{f_\pi}}$}
\put(7.5,0.5){\bf $\omega_m v$}

\put(4.0,13.6){$\frac{Z_p}{Z_m}=1.3$}
\put(7.0,11.0){$\frac{Z_p}{Z_m}=0.7$}
\put(7.0,4.6){$\frac{Z_p}{Z_m}=1.3$}
\put(4.0,2.2){$\frac{Z_p}{Z_m}=0.7$}

\put(-0.8,-4.5){
\epsfysize=22.cm
\epsffile{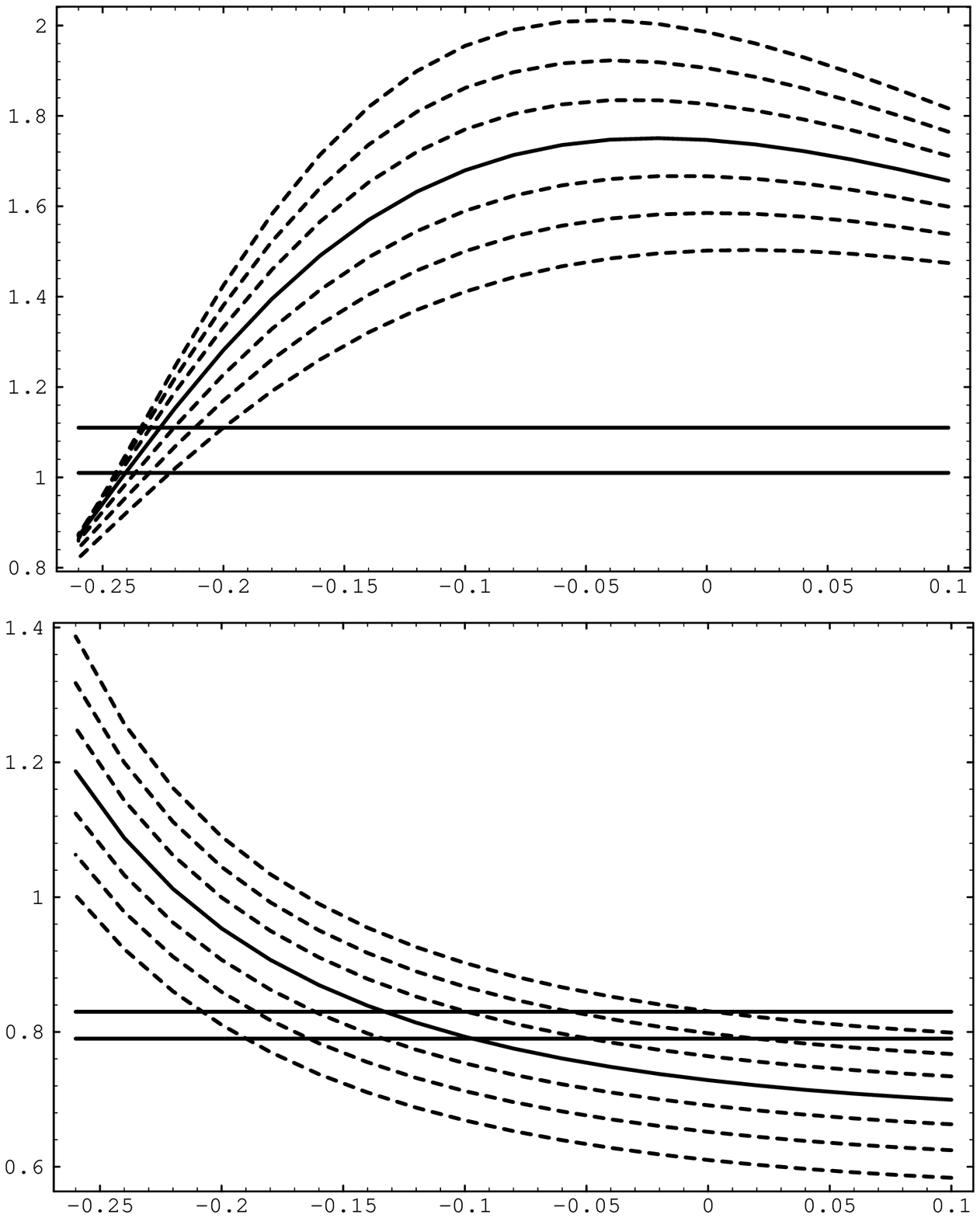}
}
\end{picture}
\caption{\footnotesize The plots show the ratios $f_\eta/f_\pi$ and
  $f_{\eta^\prime}/f_\pi$ as functions of $\omega_m v$ for various
  values of $Z_p/Z_m$.  The solid lines correspond to $Z_p/Z_m=1$ and
  the difference in $Z_p/Z_m$ between two adjacent lines is $0.1$. The
  experimentally allowed windows ($1\sigma$) for both quantities are
  bounded by the horizontal solid lines.}
\label{Plot2b} 
\end{figure}

In addition, the non--minimal kinetic terms (\ref{Lkin}) result in a
correction relevant for the $\eta$--$\eta^\prime$ mixing. This is
related to a quark mass dependence of kinetic terms and off--diagonal
kinetic terms arising from
\begin{equation}
  \label{AAA28}
  \Delta\Lc=\frac{1}{\sqrt{2}}\omega_m
  Z_m Z_h^{1/2}\tr\Phi_s\prl^\mu\Phi_p\prl_\mu\Phi_p+
  \omega_{pm}\left(Z_p Z_m Z_h\right)^{1/2}
  \prl_\mu\chi_p\tr\Phi_s\prl^\mu\Phi_p
\end{equation}
with
\begin{equation}
  \label{AAA28a}
  \ds{\omega_{pm}} = \ds{
  \frac{1}{\sqrt{6}}\left(2X_\Phi^-\ol{\sigma}_0+
  U_\Phi\right)
  \left( Z_p Z_h Z_m\right)^{-1/2}}\; .
\end{equation}
The parameter $\omega_{pm}$ induces a momentum dependence in the
$\eta$--$\eta^\prime$ mixing angle. Consequently, the mixing angle
will be different for the on--shell decays $\eta\ra2\gamma$ and
$\eta^\prime\ra2\gamma$ through which it is usually determined
experimentally. This qualitative result is, in fact, in agreement with
observation. For a quantitative comparison with experiment it is,
however, more convenient to compute the measured ratios
\begin{equation}
  \label{AAA30}
  \frac{f_\eta}{f_\pi}=\left[
  \frac{M_\eta^3}{M_\pi^3}
  \frac{\Gamma(\pi\ra2\gamma)}{\Gamma(\eta\ra2\gamma)}
  \right]^{\frac{1}{2}}\; ;\;\;\;
  \frac{f_\eta^\prime}{f_\pi}=\left[
  \frac{M_{\eta^\prime}^3}{M_\pi^3}
  \frac{\Gamma(\pi\ra2\gamma)}{\Gamma(\eta^\prime\ra2\gamma)}
  \right]^{\frac{1}{2}}
\end{equation}
from (\ref{AAA27}) and (\ref{Lkin}).  The results, as well as $M_\eta$
and $M_{\eta^\prime}$, are now functions of $U_\Phi$ and $X_\Phi^-$ or
$Z_p/Z_m$ and $\omega_m$.  Using the observed values for $M_\pi$,
$K_{K^\pm}$, $M_{K^o}$, $M_{\eta^\prime}$, $f_\pi$ and $f_{K^\pm}$ as
input we have plotted $M_\eta$ including all corrections of second
order in the quark mass expansion as a function of $\omega_m v$ for
various values of $Z_p/Z_m$ in figure~\ref{Plot1}.
Figure~\ref{Plot2b} shows the ratios (\ref{AAA30}) of decay constants
as functions of $\omega_m v$ for the same values of $Z_p/Z_m$.
The agreement with the experimental values is satisfactory for
$Z_p/Z_m\simeq0.9$ and $\omega_m v\simeq-0.19$. In addition, it is
clear from figure~\ref{Plot2b} that a phenomenologically acceptable
value of $f_\eta/f_\pi$ can only be obtained once the non--minimal
kinetic terms in (\ref{Lkin}) are taken into account. We furthermore
note that the deviation of $Z_p/Z_m$ from one is rather small. This is
compatible with the assumption that the chiral anomaly which induces a
non--vanishing $U_\vph$ is an effect which is comparable in size to
corrections linear in the quark masses.

Interestingly, the non--minimal kinetic term $\sim X_\Phi^-$ in
(\ref{Lkin}) which is responsible for the relatively large value of
$\omega_m v$ receives a large contribution from the exchange of
axial--vector mesons. More precisely, the longitudinal component of
$\prl_\mu\rho_A^\mu$, with $\rho_A$ the axial--vector meson nonet,
represents a $0^{-+}$ state which therefore mixes with the $\eta$ and
$\eta^\prime$ mesons --- a mechanism usually referred to as ``partial
Higgs effect''. A rough estimate of the contribution to $X_\Phi^-$
resulting from axial--vector meson exchange yields a value of
$\omega_m^{(\rho)} v\simeq-0.16$. This is already quite close to the
value $\omega_m v\simeq-0.19$ suggested by figures \ref{Plot1} and
\ref{Plot2b}. One is therefore tempted to even assume that {\em all}
kinetic terms (including those of $\Oc(\prl^4)$) beyond the minimal
one $\sim Z_\vph$ are dominated by the contributions from axial--vector
meson exchange. This ``leading mixing approximation'' leads, in fact,
to a satisfactory description of the $\eta$--$\eta^\prime$
system~\cite{JW96-1}.

Analogously one may apply the quark mass expansion to the scalar
sector of the linear meson model. As an interesting result one obtains
a Gell--Mann--Okubo type relation for the scalar octet
\begin{equation}
  \label{AAA31}
  M_{\sigma_\eta}^2=\frac{1}{3}\left(
  4M_{K^*}^2-M_{a_o}^2\right)
\end{equation}
where $M_{\sigma_\eta}$ denotes the mass of the scalar partner of the
$\eta$ meson. This relation can be seen to be exact to linear order in
the quark mass expansion. Using $M_{K^*}=1430\MeV$ and
$M_{a_0}=983\MeV$ as input one obtains
\begin{equation}
  \label{AAA32}
  M_{\sigma_\eta}=1550\MeV\; .
\end{equation}
Yet, it can be seen that the relation (\ref{AAA31}) is subject to
large corrections of second order in the quark mass expansion induced
by mixing of the charged scalar states with the longitudinal component
$\prl_\mu\rho_V^\mu$ of the vector meson nonet.

Finally, it should be mentioned that the linear meson model as
outlined so far also leads to satisfactory results for the ratios of
light quark masses~\cite{JW96-2}. In particular, one finds
\begin{eqnarray}
  \label{AAA33}
  \frac{m_u+m_s}{m_u+m_d}&=&\frac{M^2_{K^\pm}}{M^2_{\pi^\pm}}
  \frac{f_{K^\pm}}{f_{\pi}}\left(\frac{Z_{K^\pm}}{Z_{\pi^\pm}}
  \right)^{\frac{1}{2}}\nonumber\\
  \frac{m_u+m_s}{m_d+m_s}&=&\frac{M^2_{K^\pm}}{M^2_{K^o}}
  \frac{f_{K^\pm}}{f_{K^o}}\left(\frac{Z_{K^\pm}}{Z_{K^o}}
  \right)^{\frac{1}{2}}\; .
\end{eqnarray}
Inserting the $Z_i$ defined in (\ref{AAA29b}) with $\omega_m v$ as
determined from the $\eta$--$\eta^\prime$ sector this leads to values
for the quark mass ratios which are in excellent agreement with the
ones obtained in chiral perturbation theory. We take this as strong
support for our treatment of the linear meson model and, in
particular, for the inclusion of the non--renormalizable terms in
(\ref{AAA27}) and (\ref{Lkin}). Taking these terms into account
appears to be crucial for obtaining results for meson masses and decay
constants as well as current quark ratios in agreement with experiment
or well established results from other methods. These terms are the
main difference between our phenomenological treatment of the linear
meson model and that of previous works~\cite{Pis95-1}. 

From our phenomenological study we conclude that at least the
couplings $\ol{m}_g^2=\tr\jmath/(2N\ol{\sigma}_0)$, $\ol{\nu}$,
$\ol{\lambda}_1$ and $\ol{\lambda}_2$ should be included in the
truncation (\ref{AAA27}) of the potential. Also the difference between
$Z_h$ and $Z_m$ (which turns out to be $Z_h/Z_m\simeq0.4$ for $k\ra0$)
needs to be incorporated. This is the minimal setting for which the
meson masses have qualitatively the correct values for $k\ra0$. It
seems likely that a reasonable estimate of $\ol{\sigma}_0$ and $Z_m$
and therefore the flavor average of the decay constants,
$(2f_K+f_\pi)/3\simeq2\ol{\sigma}_0 Z_m^{1/2}$, becomes already
possible with this type of truncation. For a more detailed comparison
with experiment the coupling $X_\Phi^-$ should also be taken into
account.  A computation of $\ol{\lambda}_3$ would be helpful for an
understanding of the mass splitting in the scalar $(0^{++})$ sector.

We conclude that it would be very desirable to gain additional
information on the parameters of the linear meson model contained in
(\ref{AAA27}) and (\ref{Lkin}). This would allow to establish
relations between different scalar and pseudoscalar meson observables
beyond those based on chiral symmetry. In particular, in the scalar
sector which is not very well understood this would be quite
interesting. The application of the exact RG to the effective action
(\ref{AAA27}), (\ref{Lkin}), in fact, offers this prospect due to the
infrared fixed point structure established in section
\ref{InfraredStability}. If it is indeed realized as indicated by the
simplified analysis for two flavors outlined in this work, it should
allow for a determination of many of the couplings and wave
function renormalization constants introduced in (\ref{AAA27}) and
(\ref{Lkin}). Work in this direction is in progress.

\section*{Acknowledgments}
This work was supported by the Deutsche Forschungsgemeinschaft. We
wish to express our gratitude to the organizers of {\em QCD 96} and
the {\em Workshop on Quantum Chromodynamics: Confinement, Collisions,
  and Chaos} for providing a most stimulating environment.


\end{document}